\documentclass[journal]{./TeX/IEEEtran}
\usepackage{subfiles}
\usepackage{pifont}
\usepackage{algorithm}
\usepackage{mathtools}
\usepackage{algpseudocode}
\usepackage{pbox}
\usepackage{footmisc}
\usepackage{xr}

\ifCLASSINFOpdf
  \usepackage[pdftex]{graphicx}
\else
\fi

\usepackage[colorlinks]{hyperref}
\usepackage{multirow}
\usepackage{times}
\usepackage{amsmath,amssymb}
\usepackage{algorithm}
\usepackage{epstopdf}
\usepackage[dvipsnames]{xcolor}
\usepackage{cite}
\usepackage{bbm}
\usepackage{dsfont}

{\begin{list}               
    {$\bullet$ \hfill}{
        \setlength{\leftmargin}{\parindent}
        \setlength{\parsep}{0.04\baselineskip}
        \setlength{\itemsep}{0.5\parsep}
        \setlength{\labelwidth}{\leftmargin}
        \setlength{\labelsep}{0em}}
    }
{\end{list}}

\providecommand{\cref}[1]{Chapter~\ref{#1}}

\providecommand{\fref}[1]{Figure~\ref{#1}}

\providecommand{\R}{\ensuremath{\mathbb{R}}}

\providecommand{\bydef}{\overset{\text{def}}{=}}

\renewcommand{\vec}[1]{\ensuremath{\boldsymbol{#1}}}

\providecommand{\calF}{\mathcal{F}}

\providecommand{\mG}{\mathbf{G}}
\providecommand{\mH}{\mathbf{H}}
\providecommand{\mI}{\mathbf{I}}

\providecommand{\vh}{\mathbf{h}}

\providecommand{\vk}{\mathbf{k}}

\providecommand{\vu}{\mathbf{u}}
\providecommand{\vv}{\mathbf{v}}

\providecommand{\vx}{\mathbf{x}}
\providecommand{\vy}{\mathbf{y}}
\providecommand{\vz}{\mathbf{z}}

\providecommand{\veta}{\vec{\eta}}

\providecommand{\vone}{\mathds{1}}

\newcommand{\argmin}[1]{\mathop{\underset{#1}{\mbox{argmin}}}}
\newcommand{\argmax}[1]{\mathop{\underset{#1}{\mbox{argmax}}}}

\newcommand{\cmark}{\ding{51}}
\newcommand{\xmark}{\ding{53}}

\newcommand\Tstrut{\rule{0pt}{2.6ex}}         
\newcommand\Bstrut{\rule[-0.9ex]{0pt}{0pt}}

\begin{document}

\title{ Photon Limited Non-Blind Deblurring Using Algorithm Unrolling }

\author{Yash~Sanghvi,~\IEEEmembership{Student~Member,~IEEE}, Abhiram~Gnanasambandam,~\IEEEmembership{Student~Member,~IEEE}, and~Stanley~H.~Chan,~\IEEEmembership{Senior~Member,~IEEE}%
\thanks{Y.~Sanghvi and S.~Chan are with the School of Electrical and Computer
Engineering, Purdue University, West Lafayette, IN 47907, USA.  The work of A. Gnanasambandam was completed when he was a
graduate student at Purdue University. Email: \{ ysanghvi, stanchan\}@purdue.edu, abhiram.g94@gmail.com . The work was supported by the National Science Foundation
under Grants IIS-2133032, and ECCS-2030570.}
}
\maketitle

\begin{abstract}
Image deblurring in photon-limited conditions is ubiquitous in a variety of low-light applications such as photography, microscopy and astronomy. However, the presence of the photon shot noise due to the low illumination and/or short exposure makes the deblurring task substantially more challenging than the conventional deblurring problems. In this paper, we present an algorithm unrolling approach for the photon-limited deblurring problem by unrolling a Plug-and-Play algorithm for a fixed number of iterations. By introducing a three-operator splitting formation of the  Plug-and-Play framework, we obtain a series of differentiable steps which allows the fixed iteration unrolled network to be trained end-to-end. The proposed algorithm demonstrates significantly better image recovery compared to existing state-of-the-art deblurring approaches. We also present a new photon-limited deblurring dataset for evaluating the performance of algorithms.
\end{abstract}


\section{Introduction}

\begin{figure*}[t]
\begin{tabular}{ccc}
     \includegraphics[width=0.32\linewidth]{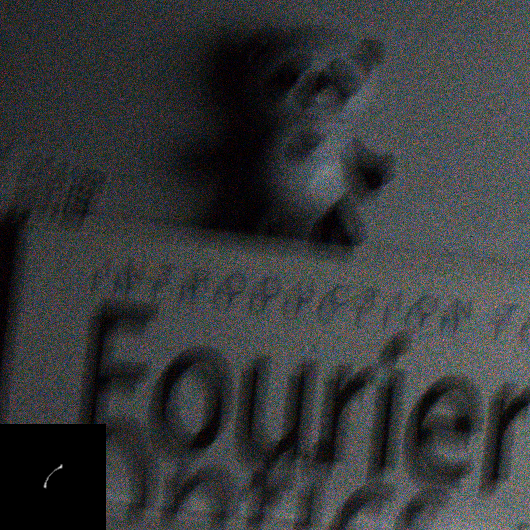}&
     \hspace{-2.0ex}\includegraphics[width=0.32\linewidth]{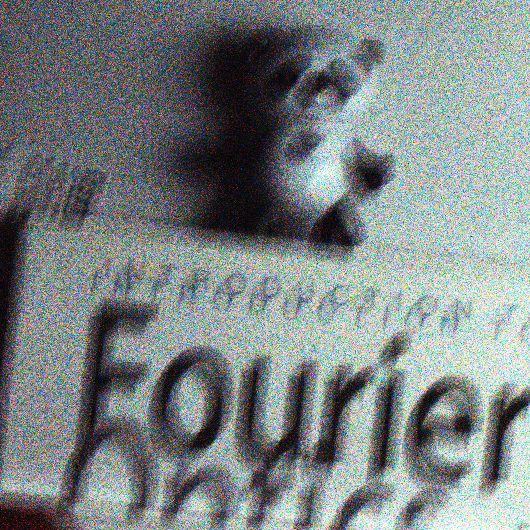}&
     \hspace{-2.0ex}\includegraphics[width=0.32\linewidth]{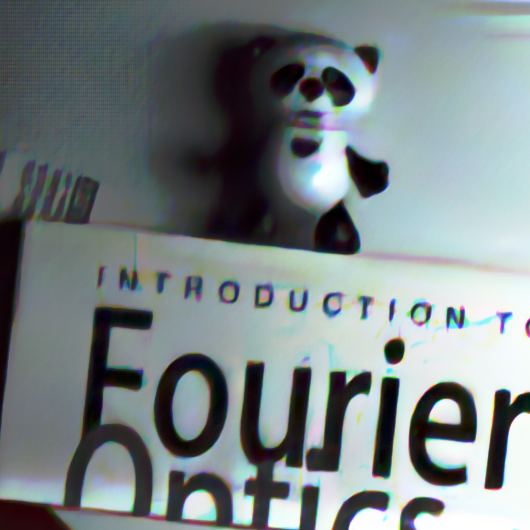}  \\
     (a) Raw camera image &\hspace{-2.0ex}(b) Histogram equalized image  &\hspace{-3.0ex}(c) Our reconstruction
\end{tabular}
\caption{\textbf{Overview.} The goal of this paper is to present a new algorithm that reconstructs images from blur at a photon-limited condition.}
\label{fig:method_demo}
\end{figure*}

\begin{figure*}
\centering
\includegraphics[page=1,width=0.99\linewidth,trim={0 120 0 0},clip]{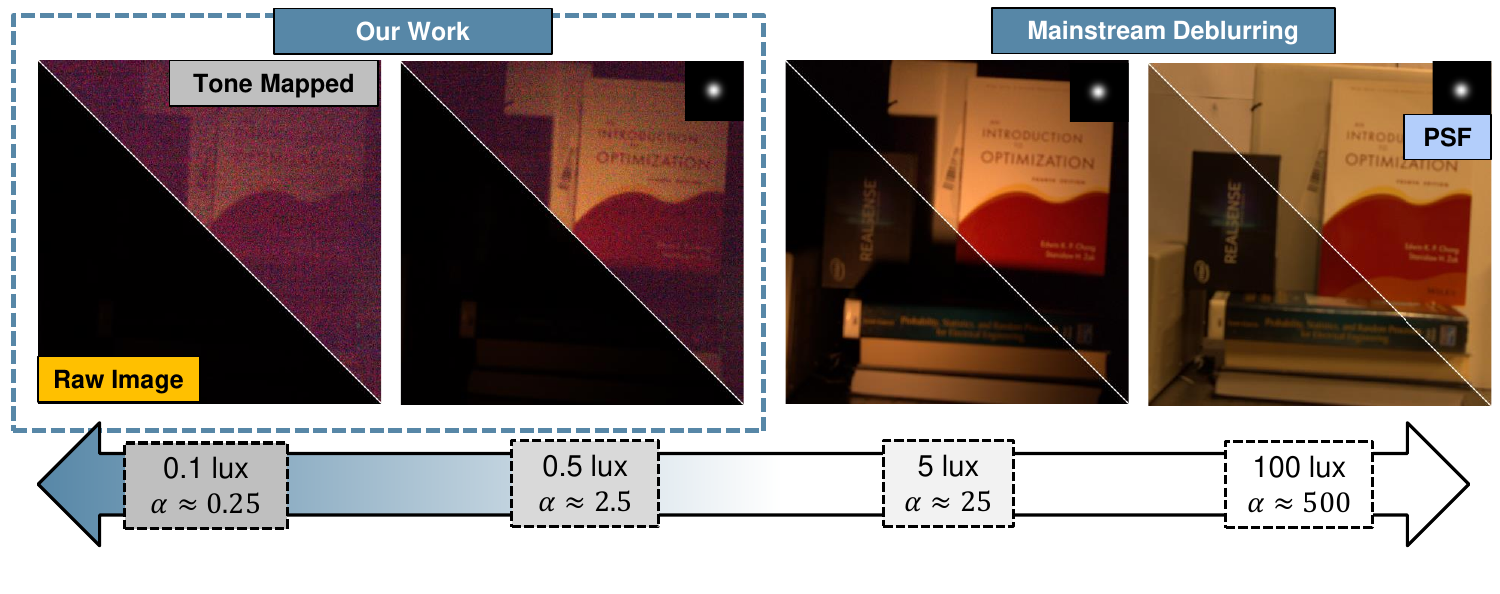}
\vspace{-2ex}
\caption{\textbf{Comparison of photon-limited scenes (Left) with relatively well illuminated scenes (Right).} Raw images and their tone mapped versions taken in different illuminations and blurred by defocus are shown in the figure. As illumination of the scene decreases, the photon shot noise becomes more dominant, making the deblurring problem substantially more difficult - as shown in \fref{fig:dwdn_fail}. In this paper, we address the problem of non-blind deblurring in a \emph{photon-limited} setting i.e. when the number of photons captured by the sensor is low leading to corruption of images by the photon shot noise. }
\label{fig:shot_noise_example}
\end{figure*}

\begin{figure}[t]
    \centering
    \includegraphics[trim={10 50 10 0},clip,page=1,width=0.8\textwidth]{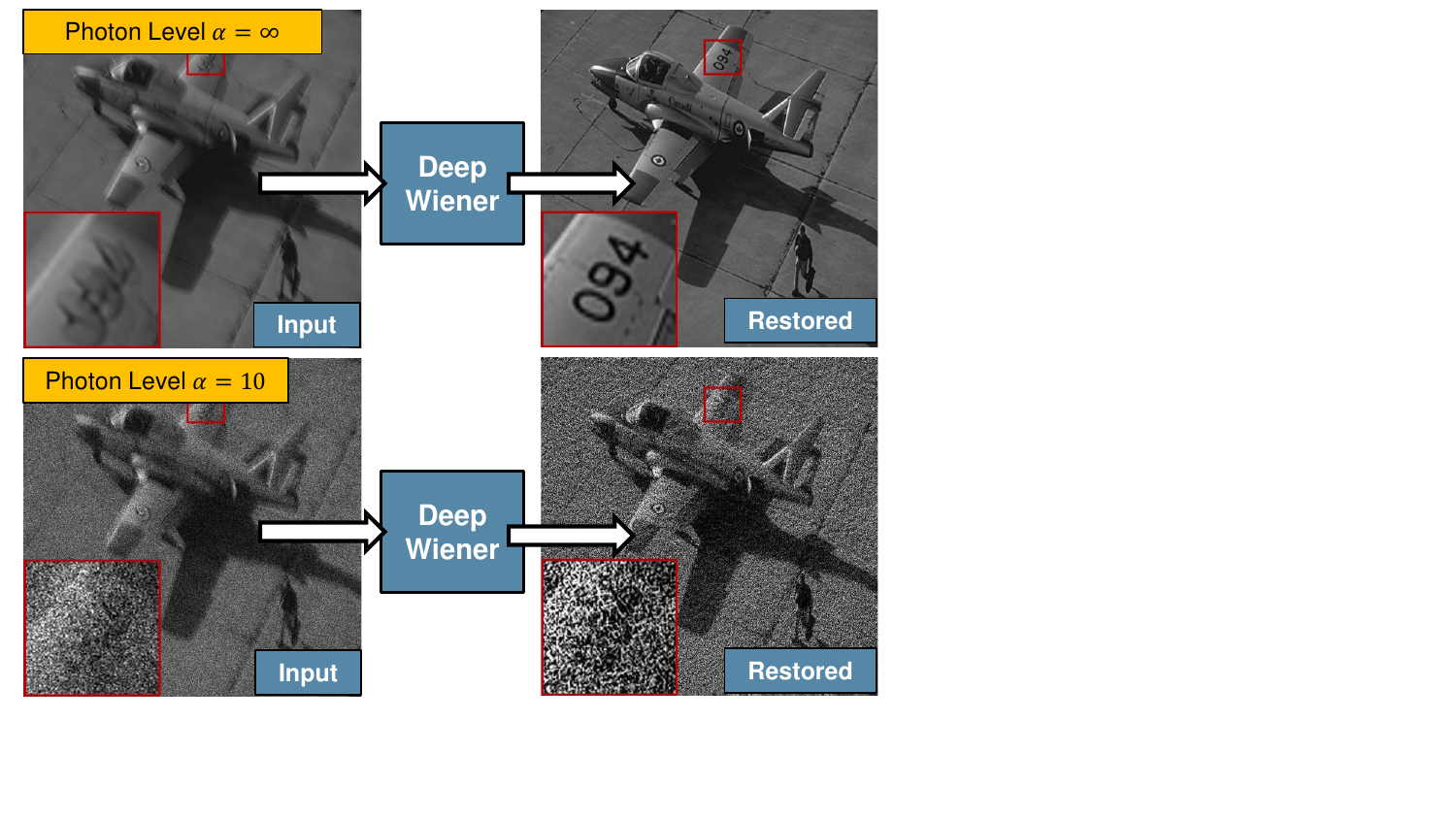}
    \caption{\textbf{Limitation of existing image deblurring algorithms when applied to low-light images.} In this example we use the pre-trained neural network \cite{dwdn} to recover a well-illuminated scene and a poorly-illuminated scene. The method fails because of the noise, even though the deblurring in a well-illuminated scene is satisfactory.}
    \label{fig:dwdn_fail}
    \vspace{-2ex}
\end{figure}

Image deblurring is a classical restoration problem where the goal is to recover a clean image from an image corrupted by a blur due to motion, camera shake, or defocus. In the simplest setting assuming a spatially invariant blur, the forward image degradation problem is
\begin{align}
    \vy = \vh \ast \vx  + \veta,
    \label{eq:non_blind_deblurring}
\end{align}
where $\vx \in \R^{N}$ is the clean image to be recovered from the corrupted image $\vy \in \R^{N}$, the vector $\vh \in R^{d}$ denotes the blur kernel, $\veta \in \R^{N}$ denotes the additive i.i.d Gaussian noise, and “$\ast$” denotes the convolution operator. The deblurring problem can be further classified as \emph{non-blind} and \emph{blind}. A non-blind deblurring problem assumes that the blur kernel $\vh$ is known whereas a blind-deblurring problem do not make such an assumption. In this paper, we focus on the non-blind case.

While non-blind deblurring methods are abundant \cite{rgdn,fdn,cpcr,nan2020variational,dwdn,dong2018denoising}, the majority are designed for well-illuminated scenes where the noise is i.i.d. Gaussian and the noise level is not too high. However, as one pushes the photon level low enough that the photon shot noise dominates, the deblurring task is no longer as simple. As illustrated in \fref{fig:method_demo}, which is a real low-light example we captured using a Canon T6i camera at a photon level approximately 5 lux, the observed image is not only dark but is strongly contaminated by photon shot noise that is visible in the histogram equalized image. To further elaborate on the operating regime of the proposed method, we show in \fref{fig:shot_noise_example} a comparison between this paper and other mainstream deblurring work. We highlight the raw sensor capture shown in the bottom left of each sub-figure and the tone-mapped image shown in the top right of each sub-figure at different illumination levels.

We refer to the problem of interest as the \emph{photon limited} non-blind deblurring. Photon limited deblurring is a common problem for a variety of applications such as microscopy~\cite{wang2013scaled} and astronomy~\cite{starck2007astronomical}. One should note that photon limited imaging is a problem even if we use a perfect sensor with zero read noise and $100\%$ quantum efficiency. The photon shot noise still exists due to the stochasticity of the photon arrival process \cite{goodman2015statistical}. Therefore, the solution presented in this paper is pan-sensor, meaning that it can be applied to the standard CCD and CMOS image sensors and the more advanced quanta image sensors (QIS) \cite{chi2020dynamic,gnanasambandam2019megapixel,li2021photon}.

\subsection{Problem formulation}
\label{sec:prob_formulation}
Consider a monochromatic image $\vx \in \R^{N}$ normalized to $[0,1]$. We write the blurred image as $\mH\vx$ where $\mH \in R^{N \times N}$ represents the blur kernel $\vh$ in the matrix form. In photon-limited conditions, the observed image is given by
\begin{align}
    \vy = \text{Poisson}(\alpha \cdot \mH \vx), \label{eq:forward_model}
\end{align}
where $\text{Poisson}(\cdot)$ denotes the Poisson process, and $\alpha$ is a scalar to be discussed. The likelihood of the observed image $\vy$ follows the Poisson probability distribution:
\begin{align}
    p(\vy|\vx; \alpha ) = \prod\limits_{j=1}^N  \frac{[\alpha\mH\vx]_j^{[\vy]_j}e^{- [\alpha\mH\vx]_j}}{ [\vy]_j !} ,
    \label{eq:poiss_dist}
\end{align}
where $[\cdot]_j$ denotes the $j$th element of a vector. The scalar $\alpha$ represents the photon level. It is a function of the sensor's properties (e.g. quantum efficiency), camera settings (exposure time, aperture), and illumination level of the scene. For a given illumination, the photon level $\alpha$ can be increased by increasing the exposure time or the aperture. To give readers a better idea of the photon level $\alpha$, we give a rough estimate of the photon flux (measured in terms of lux level) in Table~\ref{table:lux_levels} under a few typical imaging scenarios.\footnote{To estimate the photon level $\alpha$ from the photon flux level, we set the scene illumination to 1 lux (measured using a light meter) and measure the corresponding photons-per-pixel from the image sensor data captured using a Canon EOS Rebel T6i.}

\begin{table}[h]
\vspace{-3ex}
    \centering
    \caption{Lighting condition and illumination level}
    \begin{tabular}{p{4cm} p{3cm}}
        \hline
        Lighting condition & Illumination (lux) \\
        \hline Sunset & 400 \\
        Dimly-lit Street & 20-50 \\
        Moonlight & 1 \\
        $\alpha = 5$ (This paper) & 1 \\
        \hline
    \end{tabular}
    \label{table:lux_levels}
    \vspace{-2ex}
\end{table}

Under such a severe lighting condition, state-of-the-art algorithms have a hard time working. In \fref{fig:dwdn_fail} we use the deep Wiener deblurring network \cite{dwdn} to deblur the image. When the illumination is strong, the method performs well. But when the illumination is weak, the algorithm performs poorly. We remark that this observation is common for many mainstream deblurring algorithms.

\subsection{Contributions and scope}
Photon-limited non-blind deblurring is a special case of the Poisson linear inverse problem. We limit the scope to deblurring so that we can demonstrate the algorithm using real low-light data.

Existing photon-limited deblurring methods are mostly deterministic \cite{purelet,richardson1972bayesian,lucy1974iterative}. To overcome the limitation of these methods, in this paper we present a deep-learning solution. We make two contributions:
\begin{enumerate}
    \item We propose an unrolled plug-and-play (PnP \cite{venkatakrishnan2013plug,chan2016plug}) algorithm for solving the non-blind deblurring problem in \emph{photon-limited} conditions. Unlike existing work such as \cite{rond2016poisson} which uses an inner optimization to solve the Poisson proximal map, we use a three-operator splitting technique to turn all the sub-routines differentiable. This allows us to train the unrolled network end-to-end (which is previously not possible), and hence makes us the first unrolled network for Poisson deblurring.
    \item We overcome the difficulty of collecting \emph{real} photon-limited motion blur kernels and images for algorithm evaluation. A dataset containing 30 low-light images and the corresponding blur kernels are produced. We make this dataset publicly available.
\end{enumerate}

\section{Related Work}
\subsection{Poisson deconvolution}
Poisson deconvolution has been studied for decades because of its important applications \cite{deblur_review}. One of the earliest and the most cited works is perhaps the Richardson-Lucy (RL) algorithm \cite{richardson1972bayesian,lucy1974iterative}. The method assumes a known blur kernel and derives an iterative scheme which converges to the maximum-likelihood estimate (MLE) of the deconvolution problem. The RL algorithm was applied to problems such as emission tomography \cite{shepp1982maximum} and confocal microscopy \cite{dey2006richardson,laasmaa2011application}. However, since the prior is not used, the quality of reconstruction is limited.

Another class of iterative methods is based on maximum-a-posteriori (MAP) estimation by using a signal prior. For example, PIDAL-TV \cite{figueiredo2010restoration} solves a MAP cost function with the total-variation (TV) regularization using an augmented Lagrangian framework. Similarly, the sparse Poisson intensity
reconstruction algorithm (SPIRAL) \cite{harmany2011spiral} looks for sparse solutions in an orthonormal basis, whereas \cite{nowak2000statistical} solves a MAP cost function with multiscale prior using the expectation-maximization algorithm.

Shrinkage based approaches such as PURE-LET \cite{purelet} assume the deconvolution output to be a linear combination of elementary functions and minimize the expected mean squared error under a joint Poisson-Gaussian noise model. This boils down to solving a linear system of equations and has been also used to solve denoising, deblurring processes under Gaussian noise assumptions \cite{blu2007sure,xue2013multi}.

Denoising under Poisson noise conditions can be viewed as a special case of the deblurring problem. One of the widely used techniques for Poisson denoising is the variance stabilizing transforms (VST) which applies the Anscombe transform \cite{anscombe1948transformation} to stabilize the spatially varying noise variance. A standard denoising method is then used, followed by the inverse Anscombe transform. In \cite{makitalo2010optimal}, it was shown that an optimal inverse transform can outperform other standard Poisson denoising methods such as  \cite{luisier2010fast,zhang2008wavelets}. The method in \cite{azzari2016variance} provides an iterative version of the denoising via VST scheme by treating last iteration's denoised image as scaled Poisson data.

\subsection{Plug-and-play}
The Plug-and-play (PnP) framework was first introduced in \cite{venkatakrishnan2013plug} as a general purpose method to solve inverse problems by leveraging an off-the-shelf denoiser. Since then, the framework has been applied to different problems like bright field electron tomography \cite{sreehari2016plug} and magnetic resonance imaging (MRI) \cite{ahmad2019plug}. Using the same principle but with the half-quadratic splitting scheme, \cite{dpir} demonstrated the use of a single denoiser for different image restoration tasks such as super-resolution, deblurring, and inpainting. Variations of PnP have also been used for Poisson deblurring \cite{rond2016poisson,he2019plug} and  non-linear inverse problems \cite{kamilov2017plug}. A stochastic version of the scheme (PnP stochastic proximal gradient method) has been proposed for inverse problems with prohibitively large datasets \cite{online_pnp}. Using the consensus equilibrium (CE) framework  \cite{buzzard2018plug}, the scheme can be extended to fuse multiple signal and sensor models.

The convergence of the Plug-and-Play scheme has been studied in detail. For example, \cite{chan2016plug} provided a variation of the scheme which was provably convergent under the assumptions of a bounded denoiser and its performance was analysed under assumptions of a graph filter denoiser in \cite{chan2019performance}.  \cite{ryu2019plug} showed that if a denoiser satisfies certain Lipshitz conditions, the corresponding Plug-and-Play scheme can be shown to converge. Furthermore, the authors proposed real-spectral normalization as a way to impose the conditions on deep-learning based denoisers.

A closely related method which provides a framework to solve inverse problems using denoisers is REgularization by Denoising (RED) \cite{romano2017little,cohen2020regularization}. The framework poses the cost function for an inverse problem as sum of a data term and image-adaptive Laplacian regularization term. This allows the resulting iterative process to be written as a series of denoising steps. In \cite{reehorst2018regularization}, it was mentioned that for RED to be valid the denoiser needs to have a symmetric Hessian.

\subsection{Algorithm unrolling}
The difficulty of running PnP and RED is that they need to iteratively use a deep network denoiser. An alternative way to implement the algorithm was proposed by Gregor and LeCun in 2010 \cite{gregor2010learning} to unroll an iterative algorithm and train it in a supervised manner. For example, one can unroll the iterative shrinkage threshold algorithm (ISTA) for the purpose of approximating sparse codes of an image. The idea of unrolled networks has been employed in various image restoration tasks such as super-resolution \cite{usrnet}, deblurring \cite{li2019algorithm,li2020efficient}, compressive sensing \cite{yang2016deep}, and haze removal \cite{padnet2019}. For a more extensive review of algorithm unrolling, we refer the reader to \cite{monga2021algorithm}. More recently, there are new attempts to relax the fixed iteration structure of unrolling by analyzing the equilibrium of the underlying operators \cite{gilton2021deep} .

As stated in \cite{monga2021algorithm}, unrolling iterative algorithms provide multiple advantages compared to generic deep learning architectures. For example, the unrolled networks provide greater interpretability and are often parameter efficient compared to their counterparts such as the U-Net \cite{unet}. Since the networks are unrolled version of iterative algorithms, they are less susceptible to problem of overfitting.

\section{Method} \label{sec:Method}
\subsection{Algorithm unrolling}
The proposed solution for the Poisson deblurring problem is to unroll the iterative PnP algorithm. We start by deriving the PnP steps. In the ``unrolled'' version of the iterative algorithm, each iteration is treated as a computing  block. Each computing block has its own set of trainable parameters. The blocks are concatenated in series with each other. The output at the end of the last block is used as the target for a supervised loss to fine-tune the trainable parameters.

Before describing the iterative algorithm we aim to unroll, we briefly describe the underlying cost function.
Most inverse problem algorithm aim to determine the MAP estimate of the underlying signal $\vx$ by maximizing the log-posterior
\begin{align}
    \vx^* = \argmax{\vx} \Big[\log p(\vy|\vx) + \log p(\vx)\Big],
    \label{eq:map_1}
\end{align}
where $p(\vx)$ denotes the natural image prior. Plugging \eqref{eq:poiss_dist} in \eqref{eq:map_1} and taking the negative of the cost function, the maximization becomes
\begin{align}
    \vx^* = \argmin{\vx} \Big[ \alpha \vone^T\mH\vx - \vy^T\log(\alpha \mH \vx)  - \log p(\vx) \Big],
    \label{eq:poiss_map}
\end{align}
where $\vone$ represents the all-one vector. Note that the factorial term $\log \vy!$ has been dropped since it is independent of $\vx$. The prior $p(\vx)$ has not been explicitly specified yet and this issue will be addressed through the use of a denoiser in the next subsection.

\subsection{Conventional PnP for Poisson inverse problems}  \label{subsec:conven_pnp}
Now we describe how the Plug-and-Play method can be applied to the Poisson deblurring problem. We start with the alternate direction of method of multipliers (ADMM) \cite{admm} formulation  -- where we convert the unconstrained optimization problem to a constrained optimization problem by performing the variable splitting $\vx=\vz$
\begin{align}
        \{\vx^*, \vz^*\} &= \argmin{\vx , \vz} \Big[ -\log p(\vy|\vx)  - \log p(\vz) \Big],  \notag\\
        &\quad \text{ subject to $\vx=\vz$,}
        \label{eq:2_operator}
\end{align}
At the minimum of the above optimization problem, the constraint $\vx^* = \vz^*$ must be satisfied and hence the constrained optimization solution is equivalent to the unconstrained solution in \eqref{eq:poiss_map}.

The augmented Lagrangian associated with the constrained problem in \eqref{eq:2_operator}  is
\begin{align}
    \begin{split}
    \{\vx^*, \vz^*, \vu^*\} = \argmin{\vx , \vz} \Big[ \alpha \vone^T\mH\vx - \vy^T\log(\alpha \mH \vx) \\ - \log p(\vz) + \frac{\rho}{2}\|\vx-\vz+\vu\|^2 - \frac{\rho}{2}\|\vu\|^2  \Big],
    \end{split}
\end{align}
where $\vu$ denotes the scaled Lagrange multiplier corresponding to the constraint $\vx=\vz$, and $\rho$ denotes the penalty parameter. The corresponding iterative updates are:
\begin{subequations}
\begin{align}
\vx^{k+1} &= \underset{\substack{\text{Proximal operator for the negative log-likelihood}}}{\underbrace{\argmin{\vx}\; \Big[ \alpha \vone^T\mH\vx - \vy^T\log(\alpha \mH \vx)  + \frac{\rho}{2}\|\vx - \widetilde{\vx}^k\|^2 \Big]}}, \label{eq:x_sub} \\
\vz^{k+1} &= \underset{\substack{\text{Proximal operator for the negative-log-prior}}}{\underbrace{\argmin{\vz}\; \Big[-\log p(\vz) + \frac{\rho}{2}\|\vz - \widetilde{\vz}^k\|^2}\Big]},  \label{eq:z_sub}\\
\vu^{k+1} &= \vu^{k} + (\vx^{k+1} - \vz^{k+1}),
\end{align}
\end{subequations}
with $\widetilde{\vx}^k \bydef \vz^k-\vu^k$ and $\widetilde{\vz}^k \bydef \vx^k+\vu^k$. In the Plug-and-Play framework \cite{venkatakrishnan2013plug,chan2016plug}, the $\vz$ update in \eqref{eq:z_sub} is implemented by an image denoiser.

\begin{figure*}[th]
\includegraphics[page=1,width=0.99\linewidth,trim={0 200 0 0},clip]{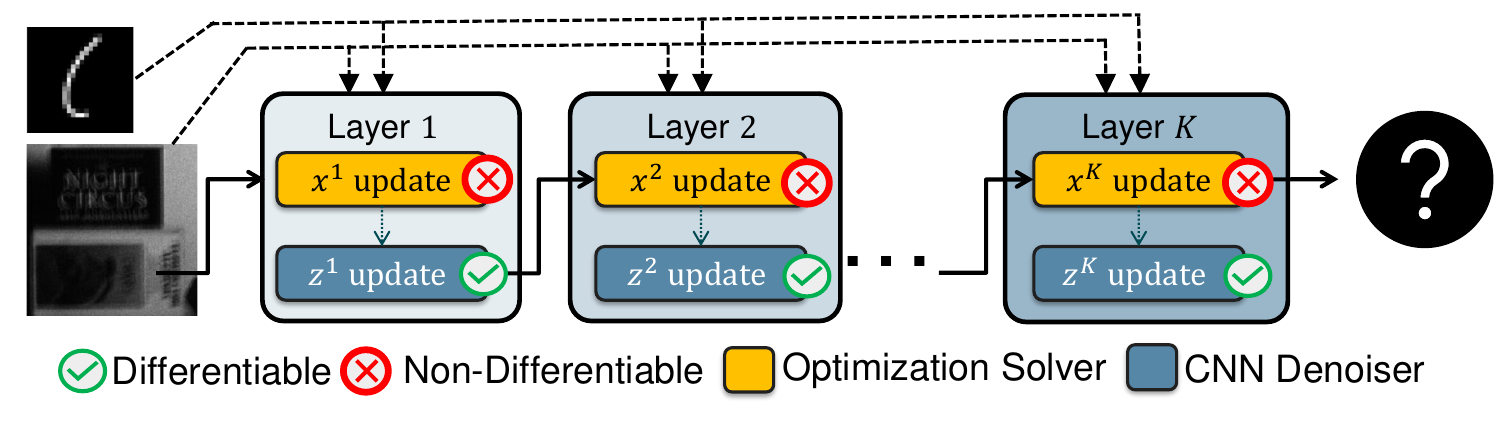}
\caption{\textbf{Conventional two-operator splitting  Plug-and-Play.} Conventional Plug-and-Play applied to the Poisson deblurring problem using equations \eqref{eq:x_sub} and \eqref{eq:z_sub}. While \eqref{eq:z_sub} is implemented as an image denoiser and hence differentiable, $\vx$-update i.e. \eqref{eq:x_sub} is implemented as a convex optimization solver and hence not differentiable. This makes the conventional PnP infeasible for fixed iteration unrolling and hence end-to-end training.}
\label{fig:2operator_splitting}
\end{figure*}

The difficulty of solving the above problem is that the $\vx$-update in \eqref{eq:x_sub} does not have a closed form expression for the Poisson likelihood. Thus \eqref{eq:x_sub} needs to be solved using an inner-loop optimization method such as L-BFGS \cite{lbfgs}. Unrolling this inner-loop optimization solver can be inefficient as it may not be differentiable. Hence unrolling the PnP scheme for the Poisson inverse problem using the existing framework is infeasible. To be more specific, while the $\vz$-update in \eqref{eq:z_sub} can be implemented as a neural network and hence is differentiable, the same cannot be said for $\vx$-update in \eqref{eq:x_sub}. As shown in \fref{fig:2operator_splitting}, when \eqref{eq:x_sub} is solved using another iterative method such as L-BFGS (for e.g. in \cite{rond2016poisson}), it is not differentiable. As a result, training the unrolled network via backpropagation is not possible unless \eqref{eq:x_sub} can be made differentiable.

\subsection{Three-operator splitting for Poisson PnP} \label{subsec:alternate_formulation}
As explained in the previous subsection, the current framework does not allow for algorithm unrolling. To circumvent this issue, we use an alternate three-operator formulation of the PnP-framework. Through this reformulation of Plug-and-Play, we derive a series of iterative updates where each step can be implemented as a single-step that is differentiable. The three-operator splitting strategy we use here has been used in context of Poisson deblurring in \cite{figueiredo2010restoration,figueiredo2009deconvolution} and \cite{he2019plug} using a TV and BM3D denoiser respectively.
\begin{figure*}[th]
    \includegraphics[page=1,width=0.95\linewidth,trim={0 50 0 0},clip]{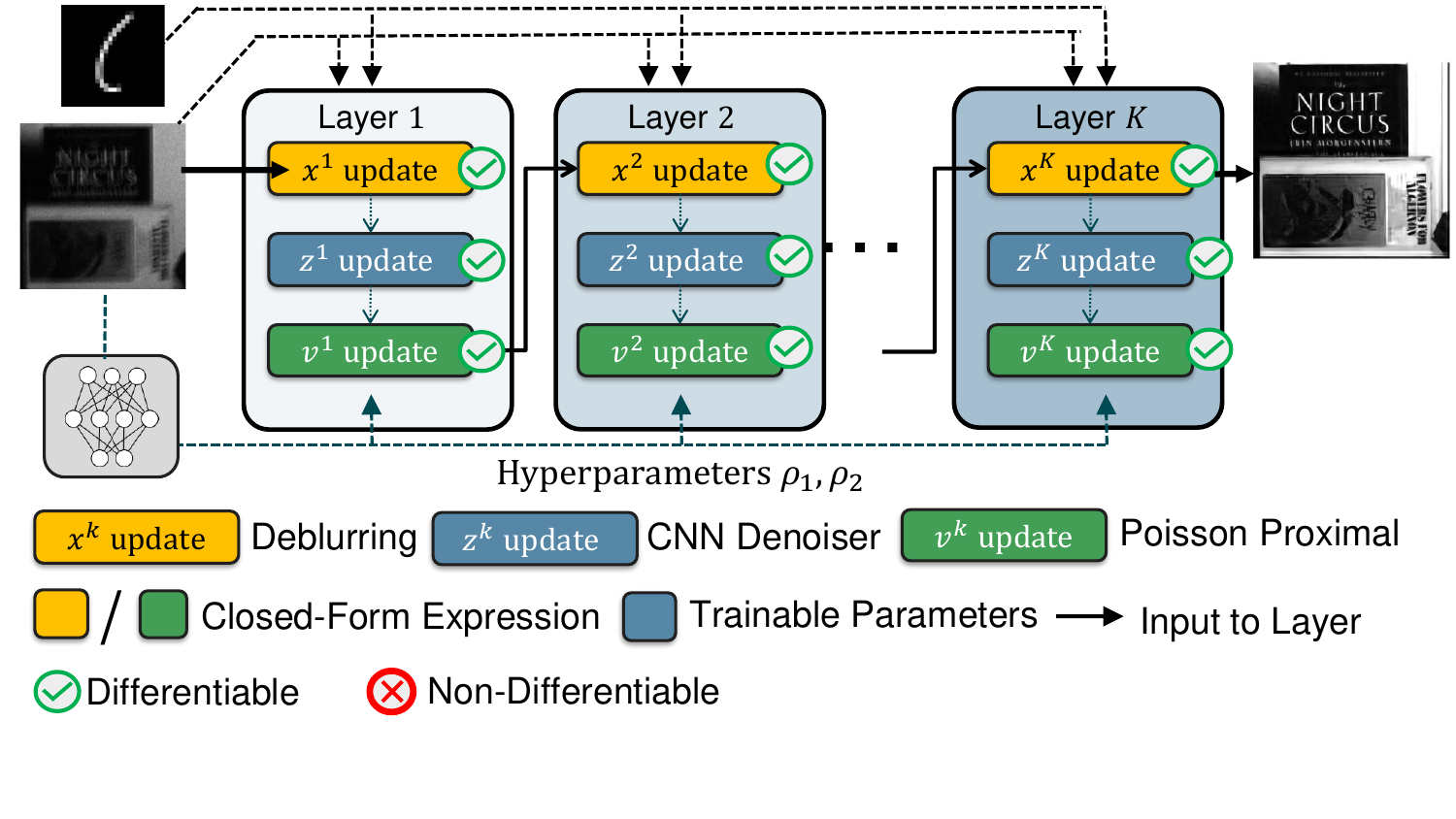}
    \caption{\textbf{Proposed unrolled Plug-and-Play for deblurring.} For conventional PnP, the data sub-problem cannot be solved in a single step and instead requires convex optimization solvers. This stops us from unrolling the iterative procedure and training it end-to-end via back-propagation. Through the three-operator splitting formulation of the problem, each sub-module in an iteration is in closed form and more importantly, differentiable. This allows for end-to-end training which was not possible in conventional PnP. The network below the input represents the hyperparameter network which predicts $\rho_1$ and $\rho_2$ using the blur kernel and the photon level.}
    \label{fig:pnp_for_poisson}
\end{figure*}

In this scheme, instead of a two-operator splitting strategy for conventional PnP in \eqref{eq:2_operator}, we use three-operator splitting to form the corresponding constrained optimization problem. Specifically, in addition to splitting the variable as $\vx = \vz$, we introduce a third variable $\vv$ corresponding to blurred image $\mH\vx$ and hence the constraint $\mH\vx = \vv$. \begin{align}
    \{\vx^*, \vz^*, \vv^* \} &= \argmin{\vx,\vz,\vv} \Big[ -\vy^T\log(\alpha\vv) + \alpha\vone^T\vv + \log p(\vz)\Big], \notag \\
    &\quad\text{subject to $\vx = \vz$, \; and \; $\mH\vx = \vv$. }
    \label{eq:pnp_3split}
\end{align}
After forming the corresponding augmented Lagrangian, we arrive at the following iterative updates:
\begingroup
\allowdisplaybreaks
\begin{subequations}
\begin{align}
    \label{eq:x_update_3split}
    \vx^{k+1} & = \argmin{\vx} \Big[ \frac{\rho_1}{2}\|\vx - \widetilde{\vx}_0^{k}\|^2 + \frac{\rho_2}{2}\|\mH\vx - \widetilde{\vx}_1^{k}\|^2 \Big], \\
    \label{eq:z_update_3split}
        \vz^{k+1} &= \argmin{\vz} \Big[ -\log p(\vz) + \frac{\rho_1}{2}\|\vz - \widetilde{\vz}_1^{k}\|^2 \Big],  \\
    \label{eq:v_update_3split}
    \vv^{k+1} &= \argmin{\vv} \Big[ -\vy^T\log(\alpha\vv) + \alpha\vone^T\vv + \frac{\rho_2}{2}\|\vv - \widetilde{\vv}^{k}\|^2 \Big],\\
    \label{eq:u1_bar_update_3split}
    \vu_1^{k+1} &=  \vu_1^{k} + \vx^{k+1} - \vz^{k+1}, \\
    \label{eq:u2_bar_update_3split}
    \vu_2^{k+1} &=  \vu_2^{k} + \mH\vx^{k+1} - \vv^{k+1},
\end{align}
\end{subequations}
\endgroup
where $\widetilde{\vx}_0^{k} \bydef \vz^{k+1} - \vu_1^{k}$, $\widetilde{x}_1^{k} \bydef \vv^{k+1} - \vu_2^{k}$, $\vv^{k} \bydef \mH\vx^{k} + \vu_2^{k}$, and $\widetilde{\vz}^{k} \bydef \vx^{k} + \vu_1^{k}$. Similar to the PnP formulation described in last subsection, the vectors $\vu_1, \vu_2$ denote the scaled Lagrangian multipliers for the constraints $\vx-\vz=0$ and $\mH\vx -\vv=0$ respectively. The scalars $\rho_1, \rho_2$ denote the corresponding penalty parameters.

Each of the subproblems defined in (\ref{eq:x_update_3split}, \ref{eq:z_update_3split}, \ref{eq:v_update_3split}) have a closed form solution and are described below:

\textbf{$\vx$-subproblem}: (\ref{eq:x_update_3split}) is a least squares minimization problem, whose solution can be explicitly given as follows:
\begin{equation}
    \label{eq:x_subproblem_exp}
    \vx^{k+1} = (\mI + (\rho_2/\rho_1)\mH^T\mH)^{-1}(\widetilde{\vx}_0^{k} + (\rho_2/\rho_1)\mH^T\widetilde{\vx}_1^{k}).
\end{equation}
Since $\mH$ represents a convolutional operator,  the operation can be performed without any matrix inversions using Fourier Transforms.
\begin{equation}
    \label{eq:x_subproblem_fft}
    \vx^{k+1} = \calF^{-1}\Big[\frac{\calF(\widetilde{\vx}_0^{k}) + (\rho_2/\rho_1)\overline{\calF(\vh)} \calF(\widetilde{\vx}_1^{k})}{1 + (\rho_2/\rho_1)|\calF(\vh)|^2} \Big],
\end{equation}
where $\calF(\cdot)$ represents the discrete Fourier transform of the image or blur kernel implemented using the Fast Fourier Transform after appropriate boundary padding. We refer to it as the \textit{deblurring operator}.

\textbf{$\vz$-subproblem}: (\ref{eq:z_update_3split}) is a proximal operator for the negative log prior term. Using the insight provided in Plug-and-Play scheme, \eqref{eq:z_update_3split} can be viewed as a denoising operation
\begin{equation}
    \label{eq:z_subproblem_exp}
    \vz^{k+1} = D(\widetilde{\vz}^{k}),
\end{equation}
where $D(\cdot)$ is any image denoiser. For end-to-end training, we require $D(\cdot)$ to be differentiable and trainable -- a property satisfied by all convolutional neural network denoisers.

\textbf{$\vv$-subproblem}: (\ref{eq:v_update_3split}) is a convex optimization problem but can be solved without an iterative procedure. Separating out each component of the vector minimization and setting the gradient equal to zero gives the following equation
\begin{equation}
    -\frac{[\vy]_i}{[\vv^{k+1}]_{i}} + \alpha + \rho_2 ([\vv^{k+1}]_i - [\widetilde{\vv}^k]_i) = 0,
    \label{eq:v_update_grad}
\end{equation}
for $i = 1, 2, \cdot\cdot\cdot, N$. Solving the resulting quadratic equation and ignoring the negative solution gives the following update step
\begin{equation}
    \label{eq:v_subproblem_exp}
    \vv^{k+1} =  \frac{ (\rho_2\widetilde{\vv}^{k}-\alpha) + \sqrt{(\rho_2\widetilde{\vv}^{k}-\alpha)^2 + 4\rho_2\vy}}{2\rho_2},
\end{equation}
Since the optimization problem in (\ref{eq:v_update_3split}) is a sum of the the negative log-likelihood for Poisson noise and a quadratic penalty term, we refer to this update as \textit{Poisson proximal operator}.

\begin{algorithm}[H]
\begin{algorithmic}[1]
\State \textbf{Input}: Blurred and Noisy Image $\vy$, kernel $\vh$, Photon level $\alpha$
\State Initialize $\vx^0$ using \eqref{eq:wiener}
\State $\vz^0 \leftarrow \vx^0$, $\vv^0 \leftarrow \vy$ $\vu^0_1 \leftarrow 0 $, $\vu^0_2 \leftarrow 0 $
    \For{$k = 1, 2, \cdot\cdot\cdot, K$}
    \State Update $\vx^k$ using Eq. (\ref{eq:x_subproblem_fft})
    \State Update $\vz^k$ using Eq. (\ref{eq:z_subproblem_exp})
    \State Update $\vv^k$ using Eq. (\ref{eq:v_subproblem_exp})
    \State $\vu^k_1 \leftarrow \vu_1^{k-1} + \vx^k - \vz^k\ $
    \State $\vu^k_2 \leftarrow \vu_2^{k-1} + \mH\vx^k - \vv^k\ $
    \EndFor
\State return $\vx^{K}$
\end{algorithmic}
\caption{Three-Operator Splitting for Poisson PnP}
\label{alg:3way_pnp_algo}
\end{algorithm}

The convergence of Algorithm~\ref{alg:3way_pnp_algo} has been derived in \cite{figueiredo2010restoration}. It was shown that as long as $\mG = [\mH^T, \mI]^T$ has a full column rank, the three-operator splitting scheme converges. Furthermore, assuming the denoiser $D$ is continuously differentiable and $\nabla D(\cdot)$ is symmetric with eigenvalues in $[0,1]$, convergence results in \cite{sreehari2016plug} show that the corresponding negative-log prior, i.e., $-\log(p(\cdot))$ is closed, proper and convex. Combined with the result from \cite{figueiredo2010restoration}, it can be shown that the three-operator PnP scheme in Algorithm~\ref{alg:3way_pnp_algo} converges.

\subsection{Unfolding the three-operator splitting}
With an end-to-end trainable iterative process, we can now describe the unfolded iterative network. The Plug-and-Play updates described in Algorithm \ref{alg:3way_pnp_algo} are now unfolded for $K = 8$ iterations and the entire differentiable pipeline is trained in a supervised manner, as summarized in \fref{fig:pnp_for_poisson}. We refer the resulting neural network architecture as \textbf{Photon-Limited Deblurring Network (\textit{PhD-Net})}.

\textbf{Initialization}: To initialize the variable $\vx^{0}$, we use the Wiener filtering step  (not to be confused with \cite{dwdn}) :
\begin{align}
    \vx^0 = \frac{1}{\alpha}\calF^{-1}\Bigg\{ \frac{  \overline{\calF(\vh)}\calF(\vy)}{1/\alpha+|\calF(\vh)|^2}\Bigg\},
    \label{eq:wiener}
\end{align}
where the constant factor $1/\alpha$ in the denominator represents the inverse of the signal-to-noise ratio of the blurred measurements. Note that this step can be derived as an $\ell_2$ regularized solution of the deconvolution problem as well.

\textbf{Hyperparameters}: The parameters used in updates (\ref{eq:x_update_3split}), (\ref{eq:v_update_3split}) -- $\rho_1, \rho_2$ are changed for each iteration and determined in one-shot by the blurring kernel $\vh$ and photon level $\alpha$ as they control the degradation of the image. The kernel $\vh$ is used as input to 4 convolutional layers, flattened to a vector of length $1024$. Along the with the photon level $\alpha$, the flattened vector is used as an input to a 3-layer fully connected network which output two set of vectors i.e. $\{\rho_1^1, \rho_1^2, ..., \rho_1^{K}\}$ and  $\{\rho_2^1, \rho_2^2, ..., \rho_2^{K}\}$. We refer the readers to the supplementary document for further architectural details.

Note that there is no ground-truth assumed for parameters $\rho_1, \rho_2$ as the hyperparameter network described above is trained simultaneously as rest of the parameters of the network.

\textbf{Denoiser}: For the denoiser used in (\ref{eq:z_subproblem_exp}), we use the architecture provided in \cite{usrnet} which introduces skip connections in a U-Net architecture known as ResUNet. Like a standard U-Net, there are four downsampling operations followed by 4 upsampling operations with skip connections between the upsampling and downsampling operators.  The denoiser weights are shared across the unrolling iterations instead of different set of weights for each iteration. For further details of the architecture we refer the readers to \cite{usrnet} or the supplementary document. Note that in our implementation of the architecture, we do not concatenate the denoiser input $\widetilde{\vz}^k$ with a noise level.

\section{Experiments}
\subsection{Training} \label{subsec:training}
We train the network described in section \ref{sec:Method} using $\ell_1$-loss function. We use images from the Flickr2K \cite{Flickr2K} dataset to train the network. The dataset contains a total of 2650 images of which we partition using a $80/20$ split for training and validation. All images are converted to gray-scale, scaled to a size of $256 \times 256$, and are blurred using motion kernels generated from \cite{motion_blur} and Gaussian blur kernels. Due to memory limits of GPU, random patches of size $128 \times 128$ were cropped and used as inputs for the network during training.

For training, a combination of $60$ motion kernels generated from \cite{motion_blur} and $10$ isotropic gaussian blur kernels with $\sigma$ varying from $\big[0.1, 2.5\big]$ were used. All the kernels were pre-generated prior to training and were randomly selected during training. Entries of the blur kernel are non-negative and sum to 1. Photon Shot noise is synthetically added to the blurred image according to \eqref{eq:forward_model}. The photon level $\alpha$ is uniformly sampled from the range $\big[1,60\big]$.

\begin{figure*}[th]
     \includegraphics[width=0.49\linewidth]{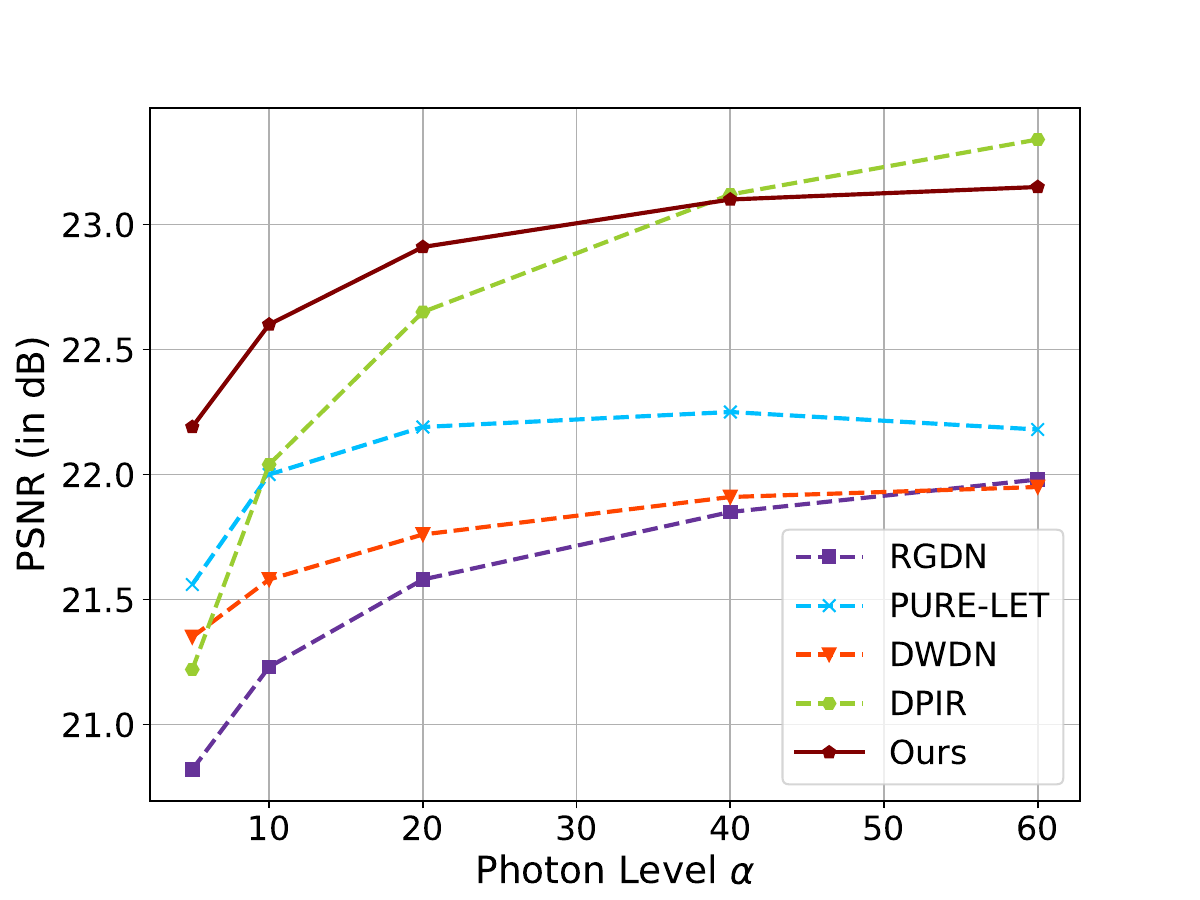}
    \includegraphics[width=0.49\linewidth]{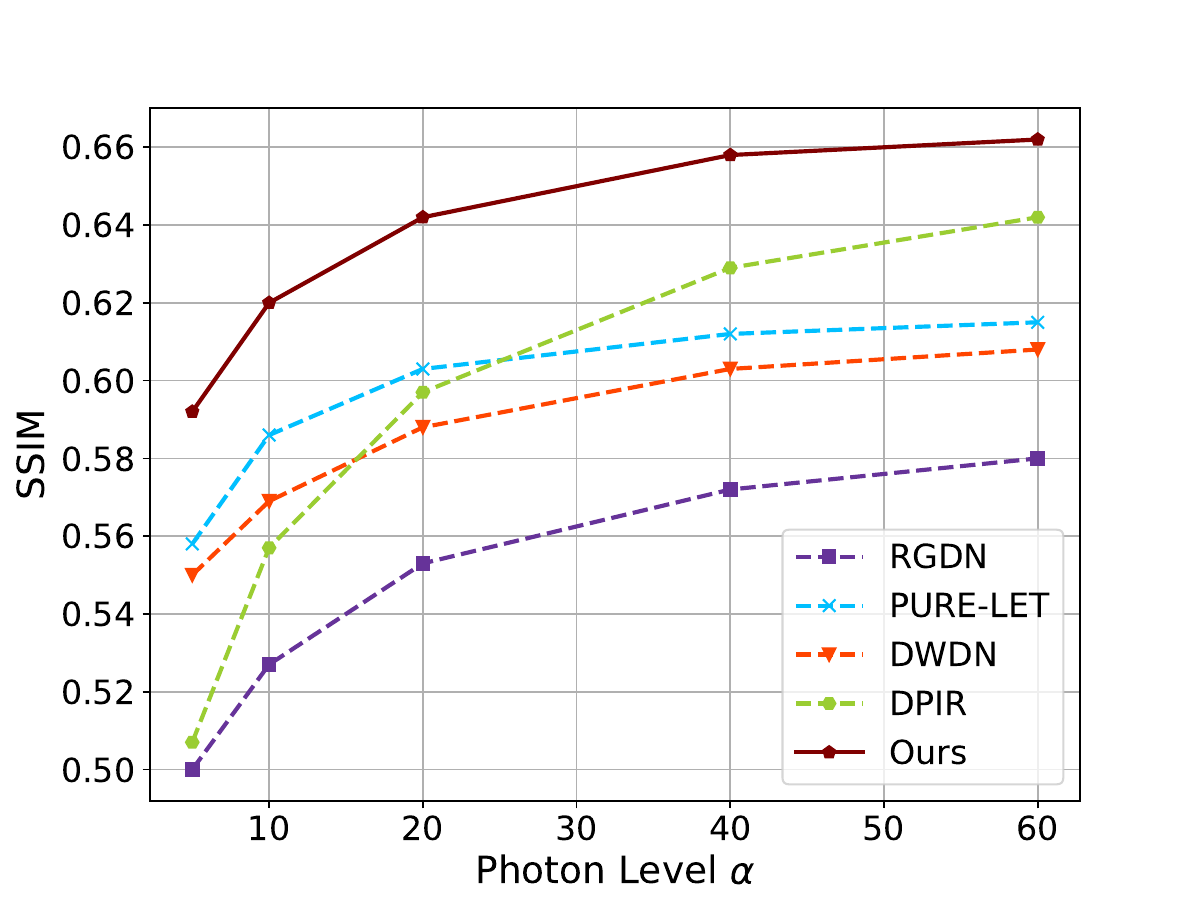}
    \caption{\textbf{Quantitative evaluation.} Comparison of PSNR and SSIM of the different methods on Levin et. al. dataset \cite{levin}. The dataset consists of 32 blurred images generated by blurring 4 images by 8 motion kernels and average PSNR/SSIM for all images and kernels plotted for different photon levels. The images were corrupted by Poisson noise at photon levels $\alpha = 5, 10, 20, 40$ and $60$.  }
    \label{fig:plot_levin}
\end{figure*}

The inputs to the network consist of the blurred and corrupted image $\vy$, the normalized blur kernel $\vh$, and the photon noise level  $\alpha$. The output from the network is the reconstructed image $\vx^{K}$ where $K$ denotes the number of iterations for which the scheme is unrolled for. We set the the number of iterations in our implementation to $K =8$. Using the $\ell_1$-loss function, we train the network with Adam optimizer \cite{kingma2014adam} using a learning rate $1 \times 10^{-4}$ and batch size of 5 for 100 epochs. All the parameters of the network are initialized using Xavier initialization \cite{xavier} and is implemented in Pytorch 1.7.0. For training, we use an NVIDIA Titan Xp GP102 GPU and it takes approximately 20 hours for training to complete.

\subsection{Choice of Deblurring Methods for Comparison}
Before describing the results of quantitative evaluation, we briefly discuss the other deblurring approaches we compare our method with. The methods, namely \textbf{RGDN} \cite{rgdn}, \textbf{DWDN} \cite{dwdn}, \textbf{DPIR} \cite{dpir}, and \textbf{PURE-LET} \cite{purelet}, were chosen because they give state-of-the-art results on the deblurring problem \textit{and} because they represent different contemporary approaches to solving the non-blind deconvolution problem.

\textbf{RGDN} (Recurring Gradient Descent Network) is an unrolled optimization method. More specifically, the authors take the deconvolution cost function $||\vy-\vk*\vx||^2 + \Omega(\vx)$ and provide a gradient descent iterative scheme for it.  The second term in the cost functions represents image prior and the corresponding gradient term $\nabla \Omega(\vx)$ is estimated using a convolutional neural network and the network, after being unrolled for fixed iterations, is trained end-to-end.

\textbf{Deep-Weiner Deconvolution (DWDN}) can be viewed as a hybrid deconvolution/denoising method. As a U-Net denoiser converts an image into a smaller feature space and then reconstructs the image using a decoder, DWDN first extracts features, performs Weiner deconvolution in that feature space, and then followed by decoding to a clean image. Through this architecture choice, they are able to perform denoising through the encoder-decoder structure but also deblur the image using Weiner deconvolution.

\textbf{DPIR (Deep Plug-and-Play Image Restortation)} uses a pre-trained denoiser in a half-quadratic splitting scheme and represents a state-of-the-art method which can be used for general purpose linear inverse problems like super-resolution and deblurring. Like our approach, it also boils down to a iterative series of denoising and deblurring steps.

\textbf{PURE-LET (Poisson Unbiased Risk Estimate - Linear Expansion of Thresholds)
} proposes the solutions as a linear combination of basis function whose weights are determined by minimizing the unbiased estimate of the mean squared loss under given noise conditions. While not a deep-learning method, it performs surprisingly competitively and can incorporate both Poisson shot noise and Gaussian read noise explicitly.

Unrolled network have received a growing interest in the signal and image processing community \cite{monga2021algorithm}. However, the vast majority of the methods are based on the Gaussian likelihood \cite{padnet2019,diamond2017unrolled}. Since our problem is Poisson, comparing our method against those Gaussian-based unrolled networks is a mismatch. Replacing the Gaussian likelihood with a Poisson likelihood would resolve this issue, but doing so would require a redesign of the unrolled network which is exactly the purpose of this paper. As such, the most relevant evaluation would be a comparison between the various two-way splitting and the three-way splitting strategies which will be shown in Section IV.D. Other unrolled methods such as \cite{li2019algorithm,li2020efficient} are designed for blind deconvolution. The work we consider here is non-blind deconvolution.

\subsection{Quantitative Evaluation}
\begin{figure}[h]
    \centering
    \begin{tabular}{c c c c}
        \multicolumn{4}{c}{Isotropic Gaussian} \\
        \includegraphics[width=0.23\linewidth]{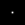} &   \hspace{-2ex}\includegraphics[width=0.23\linewidth]{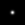} &
        \hspace{-2ex}\includegraphics[width=0.23\linewidth]{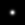} &
        \hspace{-2ex}\includegraphics[width=0.23\linewidth]{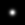} \\
        \multicolumn{4}{c}{Anisotropic Gaussian} \\
        \includegraphics[width=0.23\linewidth]{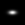} &   \hspace{-2ex}\includegraphics[width=0.23\linewidth]{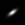} &
        \hspace{-2ex}\includegraphics[width=0.23\linewidth]{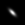} &
        \hspace{-2ex}\includegraphics[width=0.23\linewidth]{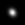} \\

        \multicolumn{4}{c}{Motion Kernel} \\
        \includegraphics[width=0.23\linewidth]{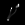} &   \hspace{-2ex}\includegraphics[width=0.23\linewidth]{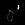} &
        \hspace{-2ex}\includegraphics[width=0.23\linewidth]{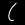} &
        \hspace{-2ex}\includegraphics[width=0.23\linewidth]{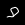} \\
    \end{tabular}
    \caption{\textbf{Kernels used for evaluation on BSD100 dataset.}}
    \label{fig:kernels}
\end{figure}

The results are summarized in \fref{fig:plot_levin}. We evaluate our method using synthetically generated noisy blurred images on 100 images from the BSDS300 dataset \cite{bsds300}, from now on referred to as \emph{BSD100}. We evaluate the performance on different photon levels ($\alpha = 5, 10, 20, 40$) representing various levels of degradation in terms of signal-to-noise ratio. We test the methods for different blur kernels - specifically 4 isotropic Gaussian kernels, 4 anisotropic Guassian kernels, and 4 motion kernels, as illustrated in \fref{fig:kernels}. Note that the top-left kernel's width is very small - this can be viewed as an identity operator and hence equivalent to evaluating the method's performance on denoising (as opposed to deblurring).

\begin{table}[ht]
    \centering
    \caption{Different features of methods used in this paper for Poisson deblurring. We classify the methods based on three criteria - iterative/non-iterative, end-to-end trainability and whether the model explicitly incorporates the fact that the images are corrupted by Poisson shot noise.}
    \begin{tabular}{p{15ex} p{10ex} p{15ex} p{15ex}}
    \hline
    \Tstrut \multirow{2}{*}{\textbf{Method}} & \multirow{2}{*}{\textbf{Iterative?}} & \textbf{End-to-End Trainable?} & \textbf{Handles Poisson Noise?}  \Bstrut\\
    \hline
    RGDN \cite{rgdn} & \cmark & \cmark & \xmark  \Tstrut\\
    PURE-LET \cite{purelet} & \xmark & \xmark & \cmark  \\
    DWDN \cite{dwdn} & \xmark & \cmark & \xmark  \\
    DPIR \cite{dpir} & \cmark & \xmark & \xmark  \\
    PhD-Net (Ours)& \cmark & \cmark & \cmark  \\
    \hline
    \end{tabular}
    \label{tab:methods}
\end{table}

As described in the previous subsection, we compare our method with the following deblurring methods - \textbf{RGDN}, \textbf{PURE-LET}, \textbf{DWDN}, and \textbf{DPIR}. Different features of the abovementioned deconvolution approaches have been summarized in Table \ref{tab:methods} for reader’s convenience.
For the sake of a fair comparison, the end-to-end trainable methods RGDN and DWDN were retrained using the same procedure as that of our method.

\begin{figure*}
\centering
    \begin{tabular}{ccccccc}
         \multirow{2}[2]{*}[19mm]{\includegraphics[width=0.25\linewidth]{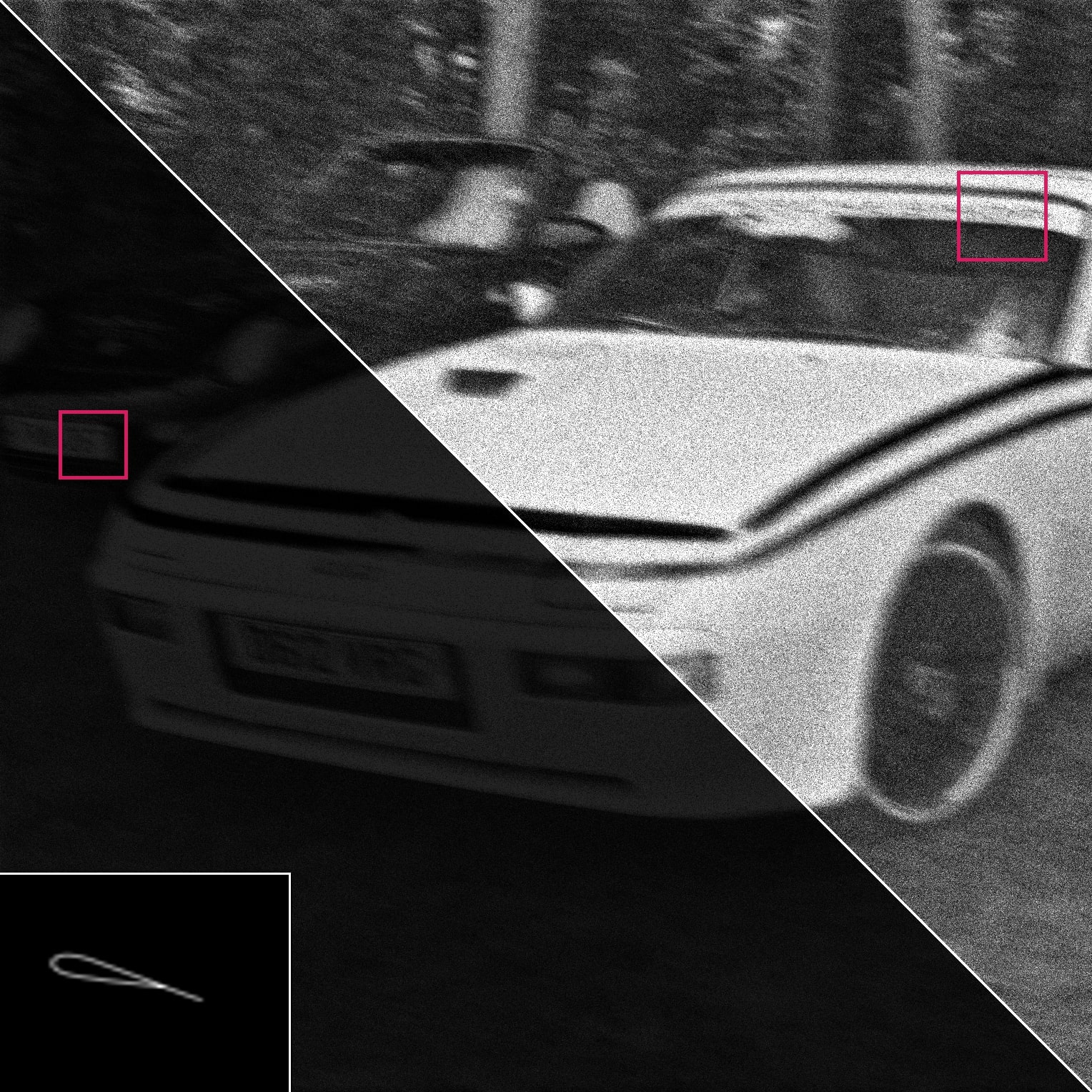}} &
         \hspace{-2.0ex}\includegraphics[width=0.12\linewidth]{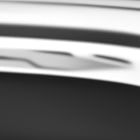}&
         \hspace{-2.0ex}\includegraphics[width=0.12\linewidth]{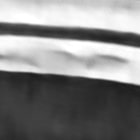}&
         \hspace{-2.0ex}\includegraphics[width=0.12\linewidth]{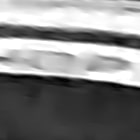}&
         \hspace{-2.0ex}\includegraphics[width=0.12\linewidth]{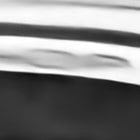}&
         \hspace{-2.0ex}\includegraphics[width=0.12\linewidth]{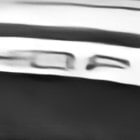}&
         \hspace{-2.0ex}\includegraphics[width=0.12\linewidth]{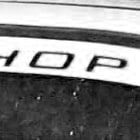}\\
         &
         \hspace{-2.0ex}\includegraphics[width=0.12\linewidth]{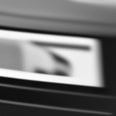}&
         \hspace{-2.0ex}\includegraphics[width=0.12\linewidth]{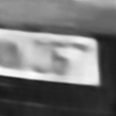}&
         \hspace{-2.0ex}\includegraphics[width=0.12\linewidth]{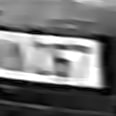}&
         \hspace{-2.0ex}\includegraphics[width=0.12\linewidth]{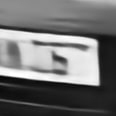}&
         \hspace{-2.0ex}\includegraphics[width=0.12\linewidth]{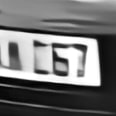}&
         \hspace{-2.0ex}\includegraphics[width=0.12\linewidth]{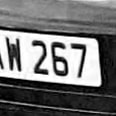}\\

          \multirow{2}[2]{*}[19mm]{\includegraphics[width=0.25\linewidth]{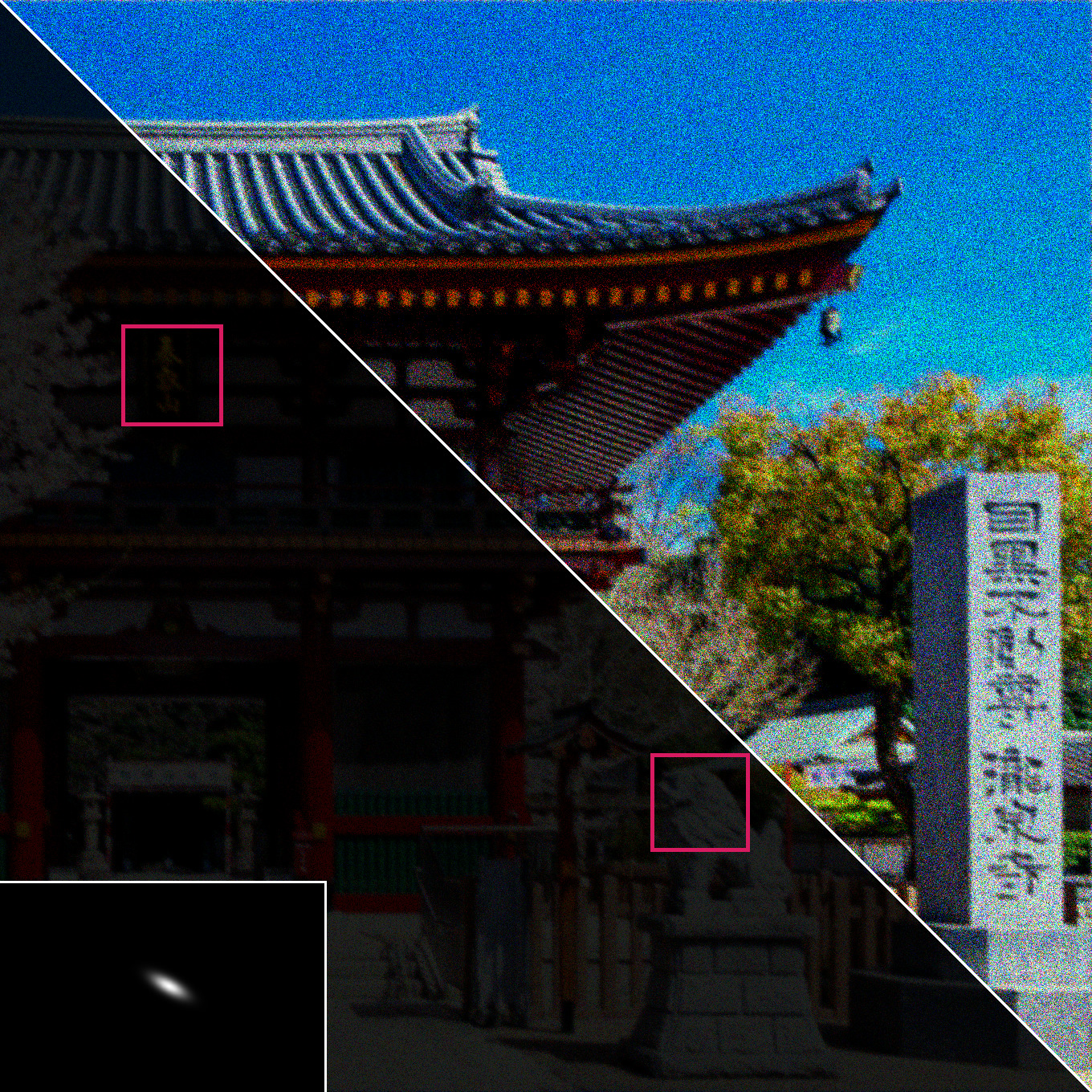}} &
         \hspace{-2.0ex}\includegraphics[width=0.12\linewidth]{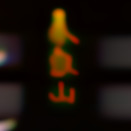}&
         \hspace{-2.0ex}\includegraphics[width=0.12\linewidth]{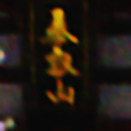}&
         \hspace{-2.0ex}\includegraphics[width=0.12\linewidth]{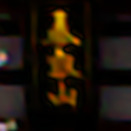}&
         \hspace{-2.0ex}\includegraphics[width=0.12\linewidth]{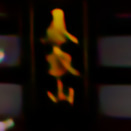}&
         \hspace{-2.0ex}\includegraphics[width=0.12\linewidth]{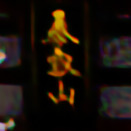}&
         \hspace{-2.0ex}\includegraphics[width=0.12\linewidth]{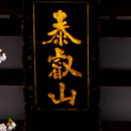}\\
         &
         \hspace{-2.0ex}\includegraphics[width=0.12\linewidth]{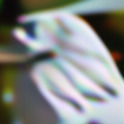}&
         \hspace{-2.0ex}\includegraphics[width=0.12\linewidth]{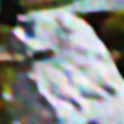}&
         \hspace{-2.0ex}\includegraphics[width=0.12\linewidth]{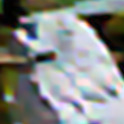}&
         \hspace{-2.0ex}\includegraphics[width=0.12\linewidth]{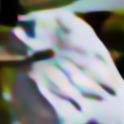}&
         \hspace{-2.0ex}\includegraphics[width=0.12\linewidth]{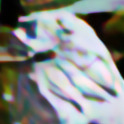}&
         \hspace{-2.0ex}\includegraphics[width=0.12\linewidth]{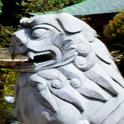}\\

         \multirow{2}[2]{*}[19mm]{\includegraphics[width=0.25\linewidth]{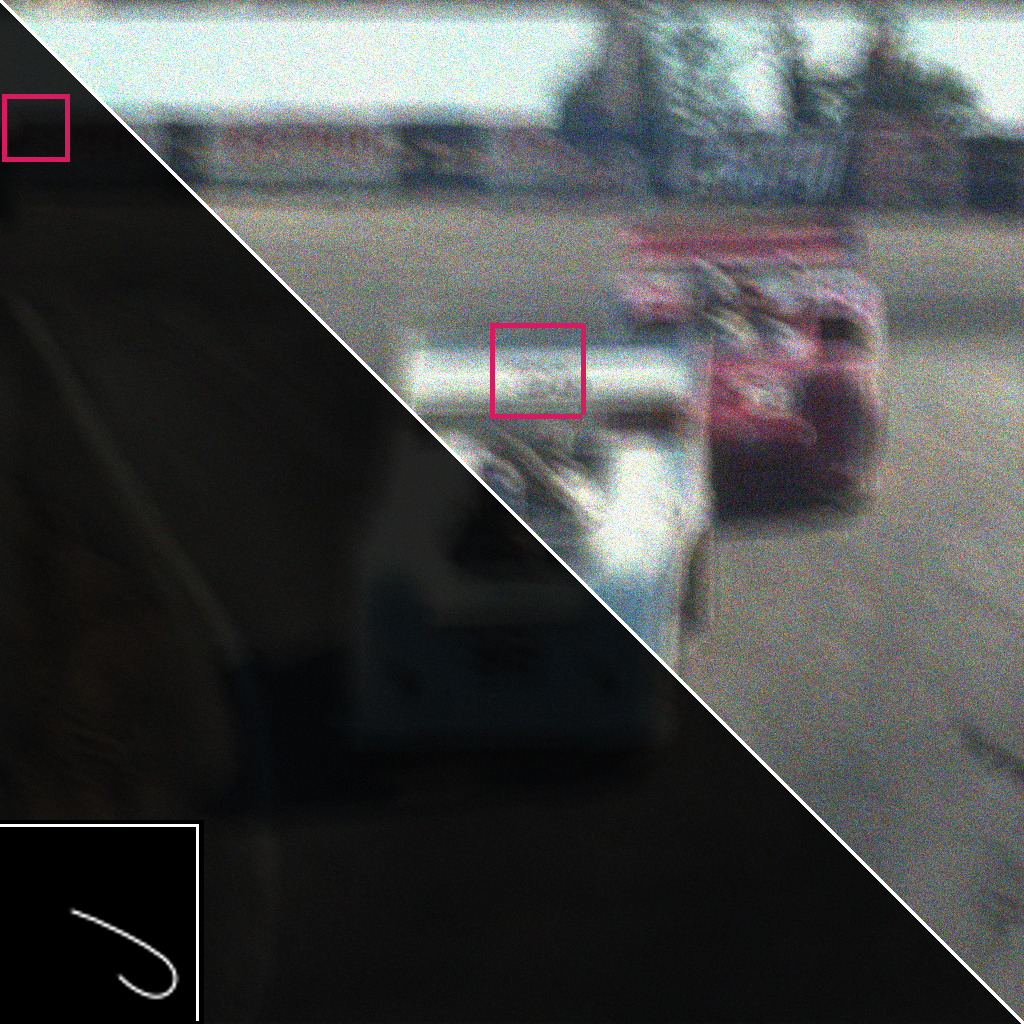}} &
         \hspace{-2.0ex}\includegraphics[width=0.12\linewidth]{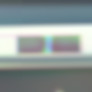}&
         \hspace{-2.0ex}\includegraphics[width=0.12\linewidth]{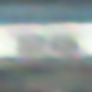}&
         \hspace{-2.0ex}\includegraphics[width=0.12\linewidth]{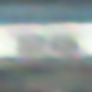}&
         \hspace{-2.0ex}\includegraphics[width=0.12\linewidth]{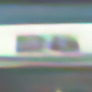}&
         \hspace{-2.0ex}\includegraphics[width=0.12\linewidth]{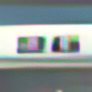}&
         \hspace{-2.0ex}\includegraphics[width=0.12\linewidth]{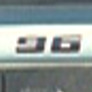}\\
         &
         \hspace{-2.0ex}\includegraphics[width=0.12\linewidth]{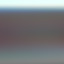}&
         \hspace{-2.0ex}\includegraphics[width=0.12\linewidth]{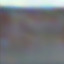}&
         \hspace{-2.0ex}\includegraphics[width=0.12\linewidth]{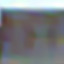}&
         \hspace{-2.0ex}\includegraphics[width=0.12\linewidth]{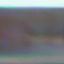}&
         \hspace{-2.0ex}\includegraphics[width=0.12\linewidth]{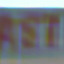}&
         \hspace{-2.0ex}\includegraphics[width=0.12\linewidth]{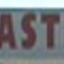}\\
         \multirow{2}[2]{*}[19mm]{\includegraphics[width=0.25\linewidth]{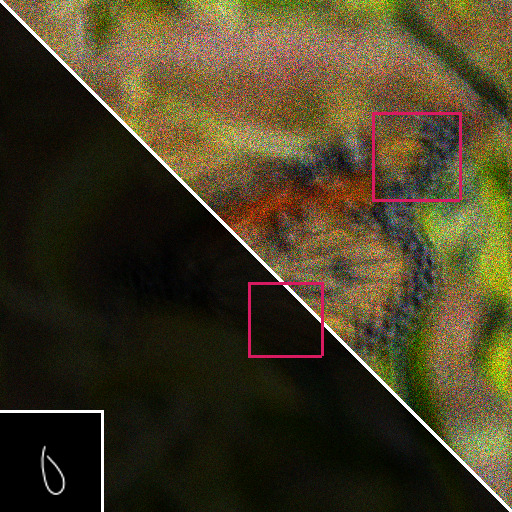}} &
         \hspace{-2.0ex}\includegraphics[width=0.12\linewidth]{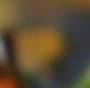}&
         \hspace{-2.0ex}\includegraphics[width=0.12\linewidth]{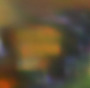}&
         \hspace{-2.0ex}\includegraphics[width=0.12\linewidth]{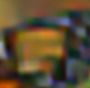}&
         \hspace{-2.0ex}\includegraphics[width=0.12\linewidth]{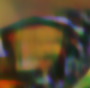}&
         \hspace{-2.0ex}\includegraphics[width=0.12\linewidth]{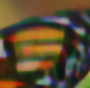}&
         \hspace{-2.0ex}\includegraphics[width=0.12\linewidth]{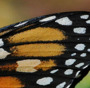}\\
         &
         \hspace{-2.0ex}\includegraphics[width=0.12\linewidth]{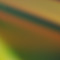}&
         \hspace{-2.0ex}\includegraphics[width=0.12\linewidth]{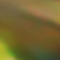}&
         \hspace{-2.0ex}\includegraphics[width=0.12\linewidth]{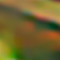}&
         \hspace{-2.0ex}\includegraphics[width=0.12\linewidth]{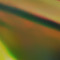}&
         \hspace{-2.0ex}\includegraphics[width=0.12\linewidth]{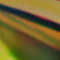}&
         \hspace{-2.0ex}\includegraphics[width=0.12\linewidth]{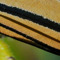}
         \\
         \multirow{2}[2]{*}[19mm]{\includegraphics[width=0.25\linewidth]{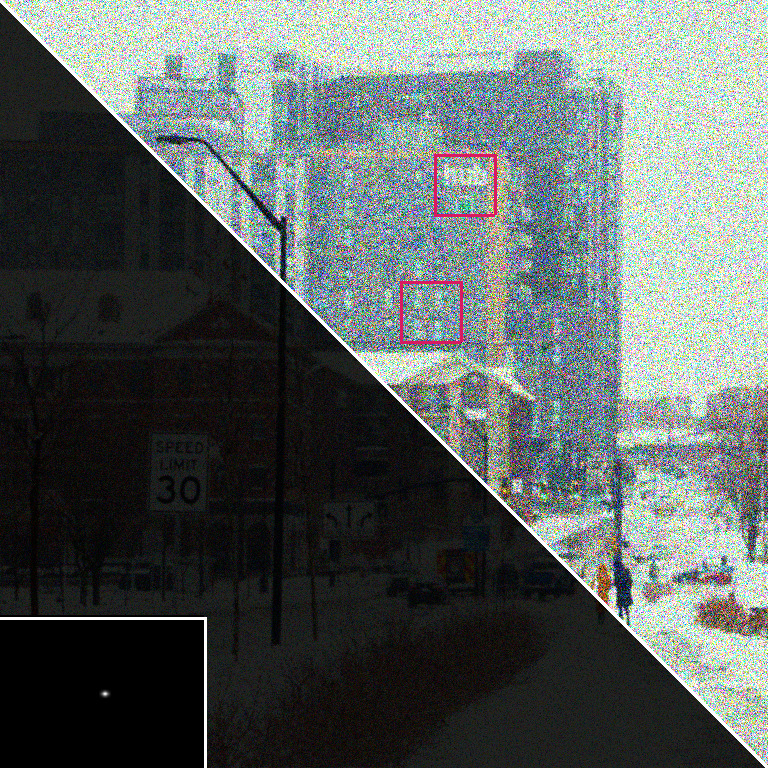}} &
         \hspace{-2.0ex}\includegraphics[width=0.12\linewidth]{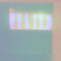}&
         \hspace{-2.0ex}\includegraphics[width=0.12\linewidth]{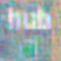}&
         \hspace{-2.0ex}\includegraphics[width=0.12\linewidth]{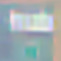}&
         \hspace{-2.0ex}\includegraphics[width=0.12\linewidth]{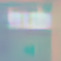}&
         \hspace{-2.0ex}\includegraphics[width=0.12\linewidth]{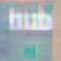}&
         \hspace{-2.0ex}\includegraphics[width=0.12\linewidth]{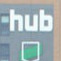}\\
         &
          \hspace{-2.0ex}\includegraphics[width=0.12\linewidth]{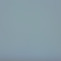}&
         \hspace{-2.0ex}\includegraphics[width=0.12\linewidth]{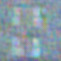}&
         \hspace{-2.0ex}\includegraphics[width=0.12\linewidth]{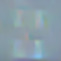}&
         \hspace{-2.0ex}\includegraphics[width=0.12\linewidth]{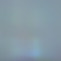}&
         \hspace{-2.0ex}\includegraphics[width=0.12\linewidth]{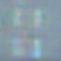}&
         \hspace{-2.0ex}\includegraphics[width=0.12\linewidth]{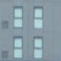}\\
         Input & \hspace{-2.0ex}DPIR \cite{dpir} & \hspace{-2.0ex}RGDN \cite{rgdn} & \hspace{-2.0ex}\small{PURE-LET} \cite{purelet} & \hspace{-2.0ex}DWDN \cite{dwdn} & \hspace{-2.0ex} PhD-Net (Ours) &\hspace{-2.0ex} Ground-Truth
    \end{tabular}
    \caption{\textbf{Qualitative Evaluation on synthetic images.} We compare the performance of the proposed method with competing methods on synthetic grayscale and color images.}
    \label{fig:qualitative_comparison}
\end{figure*}

\begin{table*}
    \caption{Comparison of proposed method with other competing approaches on BSD100 dataset}
    \resizebox{0.99\linewidth}{!}{
    \begin{tabular}{cccccccc}
        \hline
        Photon Level & Kernel & & RGDN \cite{rgdn} & PURE-LET \cite{purelet} & DWDN \cite{dwdn} & DPIR \cite{dpir} & PhD-Net (Ours)\\
        \cline{1-8}
        \multirow{6}{*}[-0.6em]{$\alpha  = 5 $} & Isotropic & PSNR (dB) & 21.77 & 22.78 & 22.50 & 22.33 & \textbf{23.46} \Tstrut\\
        & Gaussian & SSIM & 0.440 & 0.502 & 0.493 & 0.431 & \textbf{0.531} \Bstrut\\
        \cline{2-8}
        & Anisotropic & PSNR (dB) & 21.62 & 22.22 & 22.19 & 21.92 & \textbf{22.70} \Tstrut\\
        & Gaussian & SSIM & 0.427 & 0.463 & 0.464 & 0.409 & \textbf{0.491}\Bstrut\\
        \cline{2-8}
        & \multirow{2}{*}[-0em]{Motion} & PSNR (dB) & 21.14 & 21.49 & 21.54 & 21.35 & \textbf{22.12} \Tstrut\\
        & & SSIM & 0.377 & 0.419 & 0.413 & 0.377 & \textbf{0.433} \Bstrut\\
         \cline{1-8}
         \multirow{6}{*}[-0.6em]{$\alpha  = 10 $} & Isotropic & PSNR (dB) & 22.57 & 23.54 & 22.86 & 23.17 & \textbf{24.24} \Tstrut\\
        & Gaussian & SSIM & 0.491 & 0.549 & 0.527 & 0.476 & \textbf{0.576} \Bstrut\\
        \cline{2-8}
        & Anisotropic & PSNR (dB) & 22.30 & 22.81 & 22.56 & 22.60 & \textbf{23.28} \Tstrut\\
        & Gaussian & SSIM & 0.466 & 0.501 & 0.494 & 0.448 & \textbf{0.525} \Bstrut\\
        \cline{2-8}
        & \multirow{2}{*}[-0em]{Motion} & PSNR (dB) & 21.51 & 22.07 & 21.94 & 21.98 & \textbf{22.80} \Tstrut\\
        & & SSIM & 0.399 & 0.454 & 0.443 & 0.411 & \textbf{0.475} \Bstrut\\
         \cline{1-8}
         \multirow{6}{*}[-0.6em]{$\alpha  = 20 $} & Isotropic & PSNR (dB) & 23.11 & 24.27 & 23.16 & 23.98 & \textbf{24.96} \Tstrut\\
        & Gaussian & SSIM & 0.528 & 0.594 & 0.558 & 0.522 & \textbf{0.621} \Bstrut\\
        \cline{2-8}
        & Anisotropic & PSNR (dB) & 22.78 & 23.34 & 22.86 & 23.20 & \textbf{23.83} \Tstrut\\
        & Gaussian & SSIM & 0.494 & 0.536 & 0.522 & 0.485 & \textbf{0.557} \Bstrut\\
        \cline{2-8}
        & \multirow{2}{*}[-0em]{Motion} & PSNR (dB) & 21.82 & 22.70 & 22.27 & 22.65 & \textbf{23.47} \Tstrut\\
        & & SSIM & 0.418 & 0.494 & 0.475 & 0.448 & \textbf{0.515} \Bstrut\\
         \cline{1-8}
         \multirow{6}{*}[-0.6em]{$\alpha  = 40 $} & Isotropic & PSNR (dB) & 23.47 & 25.00 & 23.35 & 24.76 & \textbf{25.68} \Tstrut\\
        & Gaussian & SSIM & 0.555 & 0.638 & 0.582 & 0.569 & \textbf{0.663} \Bstrut\\
        \cline{2-8}
        & Anisotropic & PSNR (dB) & 23.10 & 23.82 & 23.10 & 23.74 & \textbf{24.36}\Tstrut\\
        & Gaussian & SSIM & 0.515 & 0.569 & 0.545 & 0.520 & \textbf{0.589}\Bstrut\\
        \cline{2-8}
        & \multirow{2}{*}[-0em]{Motion} & PSNR (dB) & 22.07 & 23.38 & 22.52 & 23.36 & \textbf{24.20} \Tstrut\\
        & & SSIM & 0.436 & 0.538 & 0.502 & 0.488 & \textbf{0.564} \Bstrut\\
        \cline{1-8}
        \end{tabular}
        }
\label{tab:psnr_list}
\end{table*}

In addition to the BSD100 dataset, we also evaluated these methods on the blurring dataset provided in Levin et. al \cite{levin}. This dataset contains a set of 32 blurred images generated by blurring 4 different clean images by 8 different motion kernels. We synthetically corrupt the blurred images with Poisson noise at different illumination levels.

The results for these evaluations are provided in Table (\ref{tab:psnr_list}) and \fref{fig:plot_levin}. For qualitative comparison on grayscale and colour reconstructions, one can refer to \fref{fig:qualitative_comparison}. On the BSDS100 dataset, our method outperforms the competing methods on all blurring kernels and illumination levels. For the dataset by Levin et. al, we outperform the other methods except DPIR at photon level $\alpha=40$. On both datasets, we observe that the gap between conventional deblurring and our method decreases as the illumination levels increase. This is because as the mean of a Poisson random variable starts increasing, the probability distribution function resembles that of a Gaussian. Therefore, the conventional deblurring methods which are designed for Gaussian noise show improved  performance.

\subsection{Comparison between 2-operator and 3-operator splitting} As explained in Section~\ref{subsec:conven_pnp}, conventional Plug-and-Play using two-operator splitting is not suitable for algorithm unrolling. The proposed three-operator splitting enables algorithm unrolling because every iterative step is differentiable. It is this end-to-end training that allows us to a better performance. In this experiment, we perform an ablation study to quantify the performance gain through different combinations of unrolling and training.

\begin{figure}[h]
    \centering
    \includegraphics[trim={15 10 25 50},clip,width=0.99\linewidth]{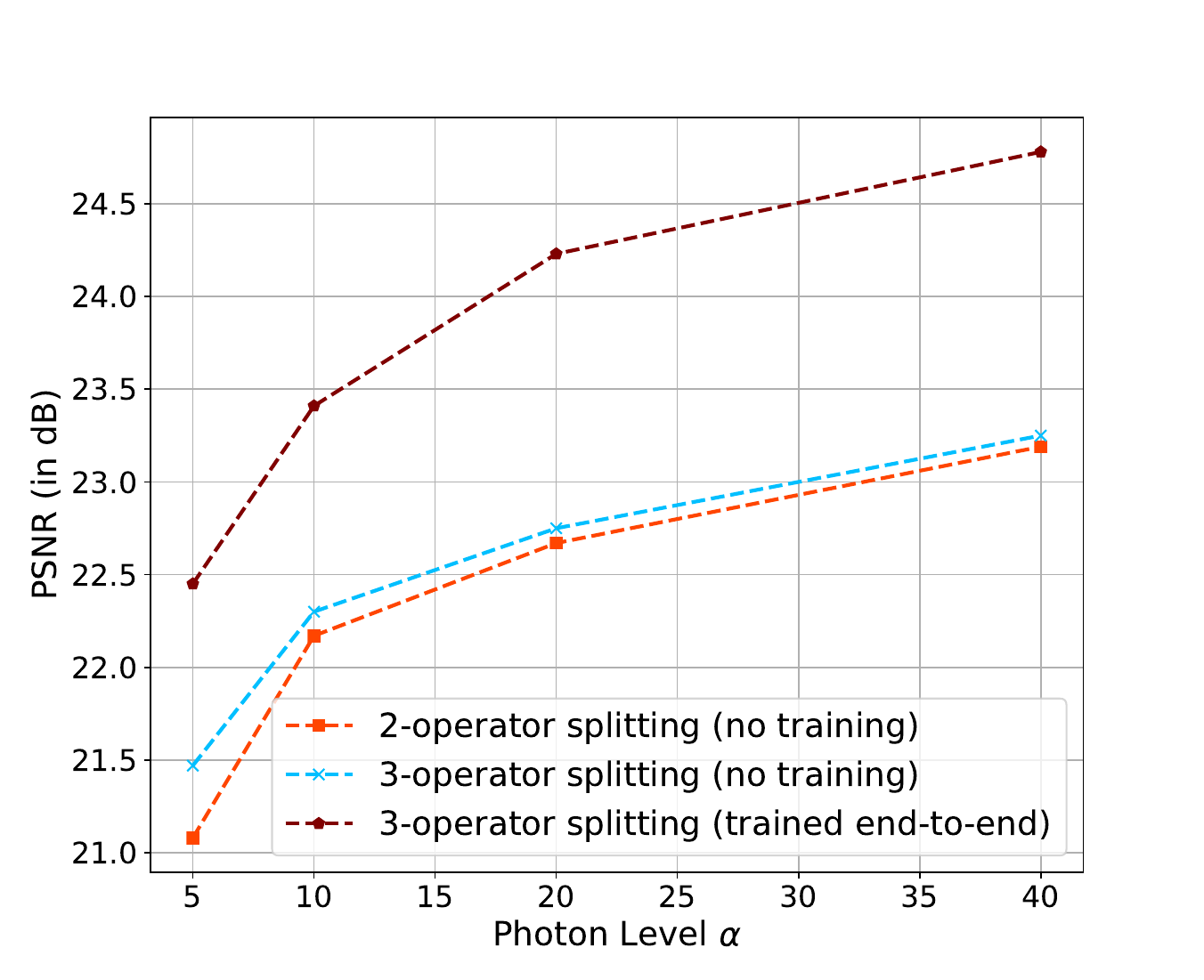}
    \caption{\textbf{Ablation study to quantify significance of algorithm unrolling.} We evaluate the following three schemes on the BSD100 dataset \textbf{(a)} conventional PnP (two-operator splitting) with a DnCNN denoiser. \textbf{(b)} alternate PnP (three-operator splitting) with a DnCNN denoiser. \textbf{(c)} proposed fixed iteration unrolled network using a DnCNN denoiser. The results of this experiment show that the significant improvement is achieved due to the network unrolling.}
    \label{fig:ablation_study}
\end{figure}

In \fref{fig:ablation_study}, we show the reconstruction performance of three schemes on the BSD100 dataset: (a) conventional two-operator splitting PnP using FFDNet denoiser as described in Section~\ref{subsec:conven_pnp} (b) an alternate three-operator splitting formulation using FFDNet as described in Section~\ref{subsec:alternate_formulation} and (c) the proposed unrolled version of the scheme described in  Section~\ref{subsec:alternate_formulation}. The results show that the two iterative schemes (a) and (b) perform similarly. However, training the proposed algorithm unrolling achieves a consistent performance gain of more than 1dB across all photon levels.

When implementing the conventional PnP in (a), we use the approach from \cite{rond2016poisson} and solve the $\vx$-update \eqref{eq:x_sub}
using a L-BFGS solver \cite{lbfgs}. Like the original implementation, we use a surrogate cost function to approximate the near zero entries with a quadratic approximation to avoid the singularities in the original cost function. A pretrained DnCNN \cite{dncnn} for noise level $\sigma=15/255$ was used for the $\vz$-update \eqref{eq:z_sub}. For the three-operator splitting scheme in (b), the same denoiser was used. To ensure a fair comparison, in the proposed fixed iteration unrolled network, we replace the ResUNet denoiser with a DnCNN and train it using the method described in Section~\ref{subsec:training}. Further details about the experiment are provided in the supplementary document.

\begin{figure*}[h]
    \centering
    \begin{tabular}{cc}
         \hspace{-3.0ex}\includegraphics[height = 6.1cm]{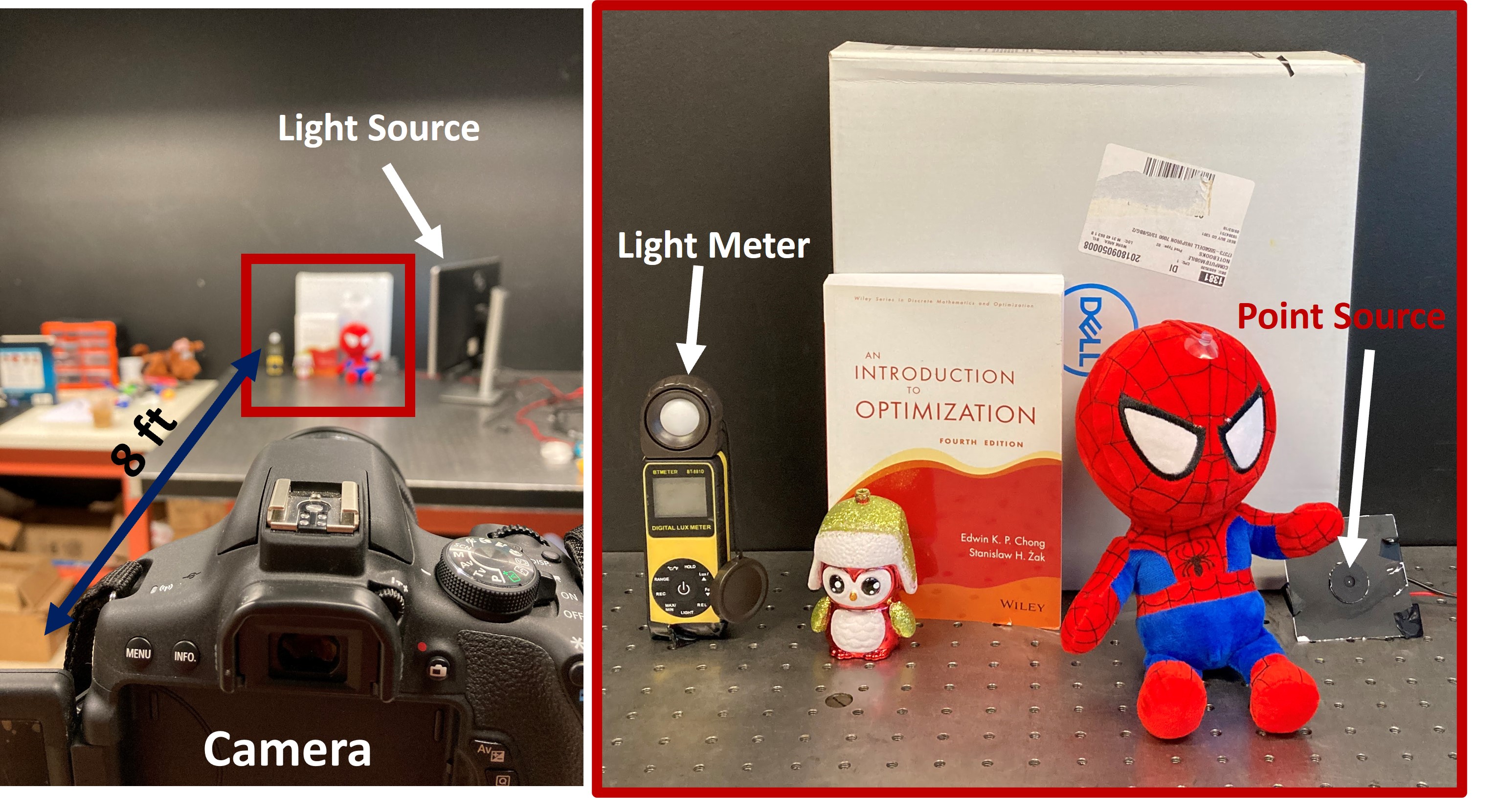}&
         \hspace{-3.0ex}\includegraphics[height = 6.1cm]{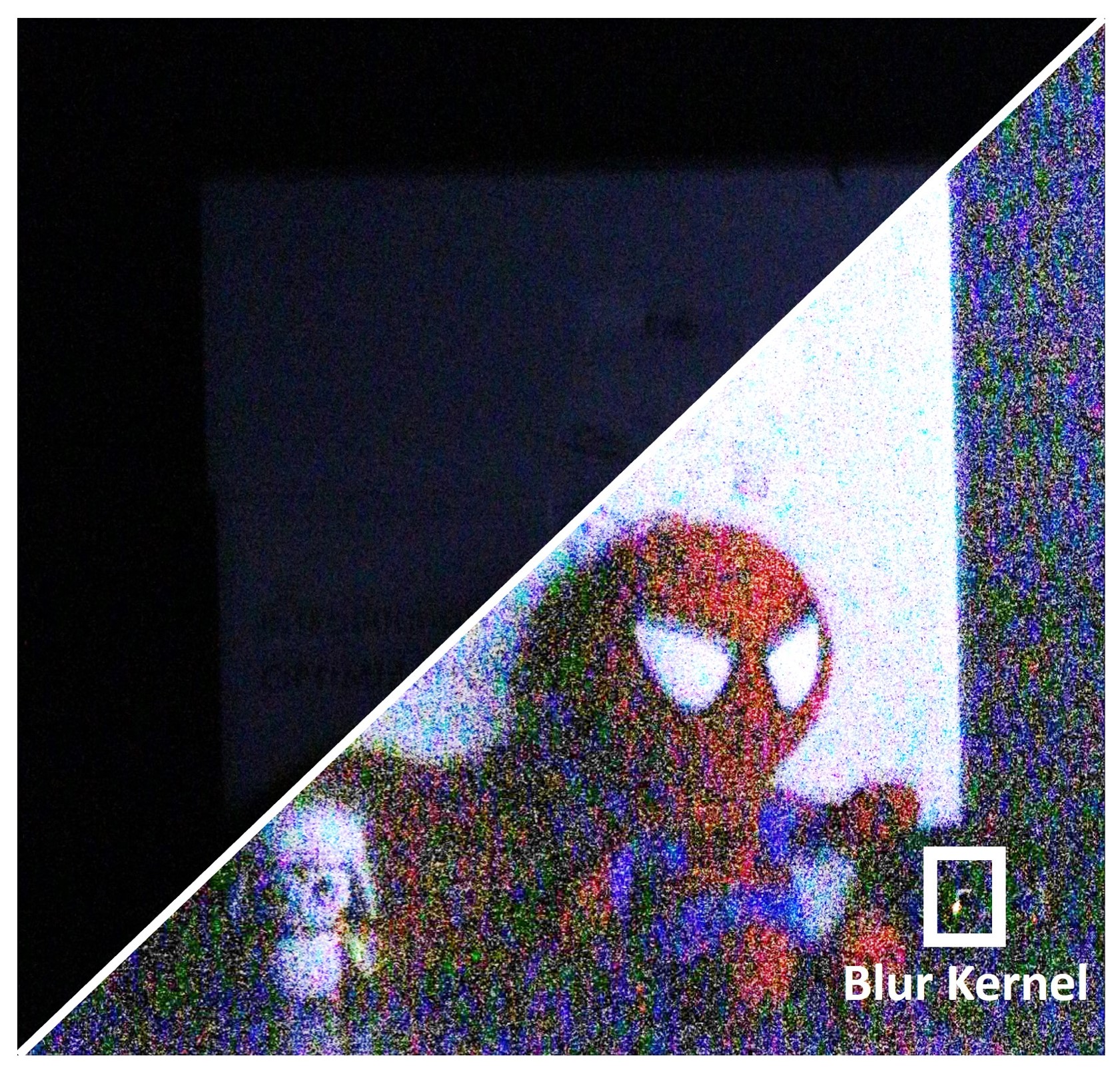}\\
         (a) Experimental setup, well illuminated scene & \hspace{-2.0ex} (b) Real capture
    \end{tabular}
    \caption{\textbf{Experimental setup}. For evaluation of the proposed method on real images, we collect noisy and blurred images using a DSLR as shown in the setup shown above. To capture a single degraded image, we reduce the illumination to a level that shot noise becomes visible. We blur image using camera shake. For the blur kernel, each scene contains a point source and the corresponding motion kernels can be visualized in \fref{fig:real_kernels}. }
    \label{fig:imaging_setup}
\end{figure*}

\subsection{Color reconstruction}
The focus of this paper is image deblurring. We acknowledge that most image sensors today acquire color images using the color filter arrays. However, adding the deblurring task with demosaicking is substantially beyond the scope of this paper. Even for demosaicking without any blur, the shot noise requires customized design, e.g., \cite{elgendy2021low}. Therefore, color images shown in this paper were processed individually for each color channel and then fused using an off-the-shelf demosaicking algorithm. While this approach is sub-optimal, our real image experiments show that the performance is acceptable.
\begin{figure}
    \begin{tabular}{ccccc}
        \includegraphics[width=0.19\linewidth]{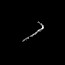} &
        \hspace{-2ex}\includegraphics[width=0.19\linewidth]{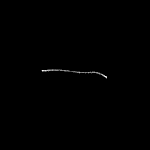}  &
        \hspace{-2ex}\includegraphics[width=0.19\linewidth]{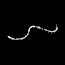} &
        \hspace{-2ex}\includegraphics[width=0.19\linewidth]{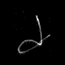} &
        \hspace{-2ex}\includegraphics[width=0.19\linewidth]{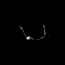} \\
        \includegraphics[width=0.19\linewidth]{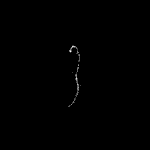} &
        \hspace{-2ex}\includegraphics[width=0.19\linewidth]{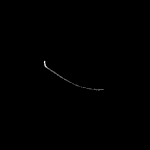}  &
        \hspace{-2ex}\includegraphics[width=0.19\linewidth]{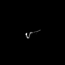} &
        \hspace{-2ex}\includegraphics[width=0.19\linewidth]{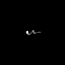} &
        \hspace{-2ex}\includegraphics[width=0.19\linewidth]{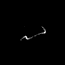}
    \end{tabular}
    \caption{ \textbf{Real kernels} generated by our optical experiment setup.}
    \label{fig:real_kernels}
\end{figure}

\section{Real Sensor Data}
Unlike conventional deblurring problems where datasets are widely available, photon-limited deblurring data is not easy to collect. In this section we report our efforts in collecting a new dataset for evaluating low-light deblurring algorithms.

\subsection{Photon-limited deblurring dataset}
We collect shot-noise corrupted and blurred images using a digital single lens reflex (DSLR) camera. The DLSR is handheld to generate motion blur. A Dell 24-inch monitor, pointing towards the region of interest, was used as a programmable illumination source to control the photon level $\alpha$. A light-meter is placed in the scene to measure the photon flux level.

\textbf{Image Capture}: We use an Canon EOS Rebel T6i camera to capture the images with exposure time of $30$ ms and aperture $f/5.0$. The ISO was set to the highest possible value of 12800 to maximize the internal gain of the sensor and hence minimize the quantization effects of the analog-to-digital convertor (ADC). The same scene was captured using different illumination levels and correspondingly different motion blur kernels. The raw image files were used for image processing instead of the compressed JPG files.

\textbf{Generating Blur}: To capture the blur kernel along with the image, we place a point source in each scene (see bottom right of middle image in  \fref{fig:imaging_setup}). The point source is created by placing an LED behind a black screen with a $30 \mu m$ pinhole. The strength of the point source is maximized to ensure the kernel is not corrupted with shot noise without saturation of pixel values. Some example kernels collected through this process can be visualized in \fref{fig:real_kernels}.

\textbf{Photon Level}: The illumination of the scenes varies between 1-5 Lux, as measured by the light-meter shown in \fref{fig:imaging_setup}. To maximize the amount of photons captured, the aperture is kept as large as possible. However shot noise is still present due to the relatively short exposure time. The estimated average photons-per-pixel (ppp) varied from $5$-$60$.

\textbf{Generating Ground Truths}: For quantitative evaluation, we also provide the ground truth for each noisy blurred image. For each noisy image corrupted by motion and noise - we place the camera on a tripod and capture 10 frames of the scene under the same illumination and camera settings. The frames, captured without any blur due to camera shake, are averaged to reduce the shot noise as much as possible. These images serve as ground truth when evaluating the performance of reconstruction methods using PSNR/SSIM.

\subsection{Reconstruction from real data}
\textbf{Pre-processing}: To reconstruct the images using our network, we first need to convert it into the format representing the number of photons captured from the raw sensor values. The raw digital data ($\vy_\text{raw}$) from the .RAW file is presented using a 14-bit value. To convert the 14-bit format to the number of photons, we use the following linear transform
\begin{align}
    \vy_{i} = \frac{\vy_{i\;\text{raw}} - b}{G},
\end{align}
where $b$ represents the zero-level offset of the camera which can be obtained from the metadata of the image .RAW file and is set equal to $b=2047$. $G$ represents the gain factor between the digital output of the sensor and the actual electrons collected by the sensor. This gain is calculated from the camera data available at \cite{photonstophotos}. Specifically, we look at the read noise of the camera in terms of digital numbers and electrons. The ratio of these two data will give the gain $G$. For Canon EOS Rebel T6i, at ISO 12800, the gain is estimated to be $G \approx 71$.

\begin{figure}[ht]
\begin{tabular}{c c}
    \includegraphics[page=1,width=0.49\linewidth,trim={0 0 340 0 },clip]{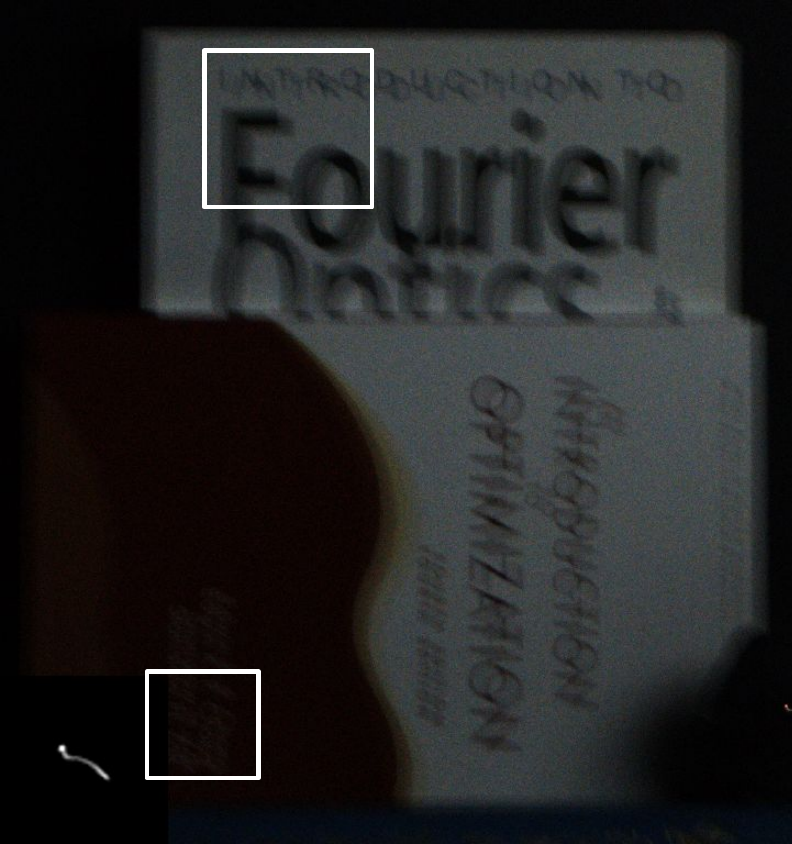} &
    \hspace{-2.0ex}\includegraphics[page=2,width=0.49\linewidth,trim={0 0 340 0 },clip]{figures_main/Figure12/Qualitative.pdf} \\
    \includegraphics[page=3,width=0.49\linewidth,trim={0 0 340 0 },clip]{figures_main/Figure12/Qualitative.pdf} &
    \hspace{-2.0ex}\includegraphics[page=4,width=0.49\linewidth,trim={0 0 340 0 },clip]{figures_main/Figure12/Qualitative.pdf} \\
    (a) Real input & (b) Processed
\end{tabular}
\caption{\textbf{Proposed method on real data.} For a qualitative comparison of other deblurring approaches on these images, refer to \fref{fig:real_data_comparison}. }
\label{fig:real_data}
\end{figure}

\begin{figure*}
\begin{tabular}{cccccc}
    \includegraphics[width=0.16\linewidth]{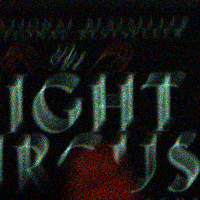} &
    \hspace{-2.0ex}\includegraphics[width=0.16\linewidth]{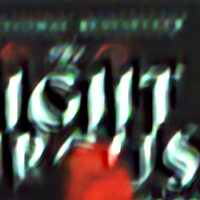} &
    \hspace{-2.0ex}\includegraphics[width=0.16\linewidth]{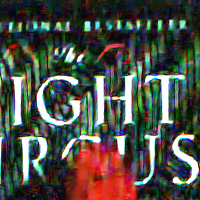} &
    \hspace{-2.0ex}\includegraphics[width=0.16\linewidth]{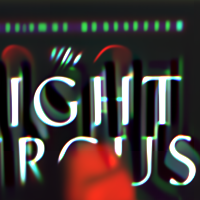} &
    \hspace{-2.0ex}\includegraphics[width=0.16\linewidth]{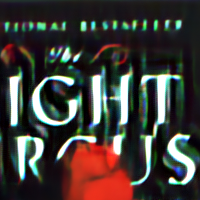} &
    \hspace{-2.0ex}\includegraphics[width=0.16\linewidth]{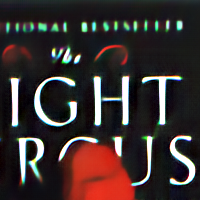}\\
    \includegraphics[angle=90,width=0.16\linewidth]{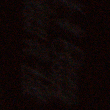} &
    \hspace{-2.0ex}\includegraphics[angle=90,width=0.16\linewidth]{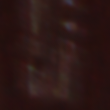} &
    \hspace{-2.0ex}\includegraphics[angle=90,width=0.16\linewidth]{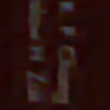} &
    \hspace{-2.0ex}\includegraphics[angle=90,width=0.16\linewidth]{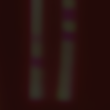} &
    \hspace{-2.0ex}\includegraphics[angle=90,width=0.16\linewidth]{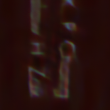} &
    \hspace{-2.0ex}\includegraphics[angle=90,width=0.16\linewidth]{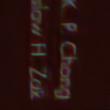} \\
    \includegraphics[width=0.16\linewidth]{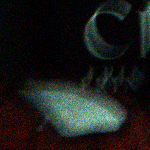} &
    \hspace{-2.0ex}\includegraphics[width=0.16\linewidth]{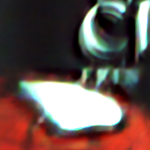} &
    \hspace{-2.0ex}\includegraphics[width=0.16\linewidth]{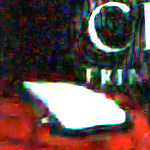} &
    \hspace{-2.0ex}\includegraphics[width=0.16\linewidth]{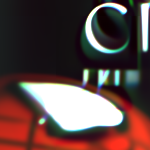} &
    \hspace{-2.0ex}\includegraphics[width=0.16\linewidth]{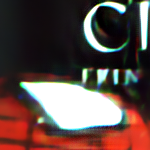} &
    \hspace{-2.0ex}\includegraphics[width=0.16\linewidth]{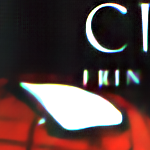}\\
    \includegraphics[width=0.16\linewidth]{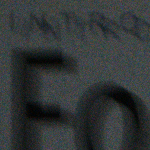} &
    \hspace{-2.0ex}\includegraphics[width=0.16\linewidth]{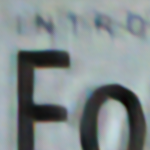} &
    \hspace{-2.0ex}\includegraphics[width=0.16\linewidth]{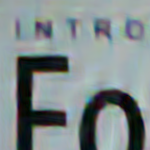} &
    \hspace{-2.0ex}\includegraphics[width=0.16\linewidth]{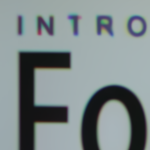} &
    \hspace{-2.0ex}\includegraphics[width=0.16\linewidth]{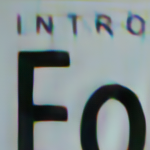} &
    \hspace{-2.0ex}\includegraphics[width=0.16\linewidth]{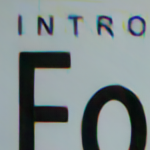}\\

    Raw Image & \hspace{-2.0ex}RGDN \cite{rgdn} & \hspace{-2.0ex}PURE-LET \cite{purelet} & \hspace{-2.0ex}DPIR \cite{dpir} & \hspace{-2.0ex}DWDN \cite{dwdn} & \hspace{-2.0ex}\textbf{PhD-Net (Ours) } \\
    PSNR (dB) / SSIM & \hspace{-2.0ex} 17.43 / 0.423 & \hspace{-2.0ex} 20.83 / 0.478 & \hspace{-2.0ex} 21.50 / 0.546 & \hspace{-2.0ex} 21.94 / 0.575 & \hspace{-2.0ex} \textbf{22.69 / 0.613}
\end{tabular}
\caption{\textbf{Qualitative Comparison} We look at zoomed in regions of the reconstructed images from \fref{fig:real_data} using competing methods. The average PSNR and SSIM evaluated on the given patches is provided at the bottom. From visual inspection one can see that our method is able to recover finer details compared to other methods. Note that in the first row, the DPIR output may look qualitatively similar to our result. This is because the former often tends to blur out images for a ``cleaner`` looking image as observed in the second row of zoomed in reconstructions. }
\label{fig:real_data_comparison}
\end{figure*}

Our reconstruction results are shown in \fref{fig:real_data}. We also compare reconstructions using proposed method with other contemporary deblurring methods (RGDN, PURE-LET, DPIR and DWDN) in \fref{fig:real_data_comparison}. Through a visual inspection, one can conclude that our method is able to reconstruct finer details from the noisy and blurred image while leaving behind fewer artifacts.

\textbf{Quantitative Evaluation}: For evaluation of metrics such as PSNR and SSIM, we register the ground truth to the reconstruction using homography transformation to account for the differences in camera positions. The average PSNR and SSIM on the real datset for the proposed and competing methods are reported in Table~\ref{tab:real_eval}. We outperform the second-best competing methods, i.e. \cite{dwdn}, by $0.6$dB in terms of PSNR and by $0.005$ in terms of SSIM. As shown in \fref{fig:real_data_comparison}, when evaluating SSIM on a few patches containing text, the gap between our method and \cite{dwdn} becomes much wider.

\begin{table}[th]
    \caption{PSNR (in dB) and SSIM evaluated on real dataset of 30 images.}
    \centering
    \resizebox{0.99\linewidth}{!}{
    \begin{tabular}{cccccc}
        \hline
        Method & RGDN \cite{rgdn} & PURE-LET \cite{purelet} & DPIR \cite{dpir} & DWDN \cite{dwdn} &  \textbf{PhD-Net (Ours)} \\
        \hline
        PSNR & 19.80 & 20.88 & 22.09 & 22.85 &  \textbf{23.48} \\
        SSIM & 0.476 & 0.501 & 0.548 & 0.561 &  \textbf{0.566} \\
        \hline
    \end{tabular}}
    \label{tab:real_eval}
\end{table}

\section{Conclusion and Future Work}
In this paper, we formulated the photon-limited deblurring problem as a Poisson inverse problem. We presented an end-to-end trainable solution using a algorithm unrolling technique. We performed extensive numerical experiments to compare our approach with other existing state-of-the-art non-blind deblurring approaches and demonstrated how our method can be applied to real sensor data. Even though the present solution is focused on image deblurring, it can be easily extended to other photon-limited inverse problems such as compressive sensing, lensless imaging, and super-resolution.

The algorithm presented in this paper can be used  to reconstruct a single clean image from multiple blurred images. This would allow us to take advantage of the temporal redundancy which would be necessary to obtain a meaningful clean signal in much challenging scenarios (e.g. photon level $\alpha \leq 5$). Another interesting but challenging problem which can be attempted using the framework is  \textit{low-light blind deconvolution} i.e. recovering the clean image and blur kernel simultaneously from blurred images corrupted with photon shot noise.

\typeout{}
\bibliography{ref}

\begin{thebibliography}{10}

\bibitem{dwdn}
J.~Dong, S.~Roth, and B.~Schiele, ``Deep wiener deconvolution: Wiener meets
  deep learning for image deblurring,'' {\em Advances in Neural Information
  Processing Systems}, vol.~33, pp.~1048--1059, 2020.

\bibitem{rgdn}
D.~{Gong}, Z.~{Zhang}, Q.~{Shi}, A.~{van den Hengel}, C.~{Shen}, and
  Y.~{Zhang}, ``Learning deep gradient descent optimization for image
  deconvolution,'' {\em IEEE Transactions on Neural Networks and Learning
  Systems}, vol.~31, no.~12, pp.~5468--5482, 2020.

\bibitem{fdn}
J.~Kruse, C.~Rother, and U.~Schmidt, ``Learning to push the limits of efficient
  {FFT}-based image deconvolution,'' in {\em Proceedings of the IEEE
  International Conference on Computer Vision}, pp.~4586--4594, 2017.

\bibitem{cpcr}
T.~Eboli, J.~Sun, and J.~Ponce, ``End-to-end interpretable learning of
  non-blind image deblurring,'' in {\em Proceedings of 16th European Conference
  on Computer Vision, Part XVII 16}, pp.~314--331, Springer, 2020.

\bibitem{nan2020variational}
Y.~Nan, Y.~Quan, and H.~Ji, ``Variational-{EM}-based deep learning for
  noise-blind image deblurring,'' in {\em Proceedings of the IEEE/CVF
  Conference on Computer Vision and Pattern Recognition}, pp.~3626--3635, 2020.

\bibitem{dong2018denoising}
W.~Dong, P.~Wang, W.~Yin, G.~Shi, F.~Wu, and X.~Lu, ``Denoising prior driven
  deep neural network for image restoration,'' {\em IEEE Transactions on
  Pattern Analysis and Machine Intelligence}, vol.~41, no.~10, pp.~2305--2318,
  2018.

\bibitem{wang2013scaled}
H.~Wang and P.~C. Miller, ``Scaled heavy-ball acceleration of the
  richardson-lucy algorithm for 3d microscopy image restoration,'' {\em IEEE
  Transactions on Image Processing}, vol.~23, no.~2, pp.~848--854, 2013.

\bibitem{starck2007astronomical}
J.-L. Starck and F.~Murtagh, {\em Astronomical Image and Data Analysis}.
\newblock Springer Science \& Business Media, 2007.

\bibitem{goodman2015statistical}
J.~W. Goodman, {\em Statistical Optics}.
\newblock John Wiley \& Sons, 2015.

\bibitem{chi2020dynamic}
Y.~Chi, A.~Gnanasambandam, V.~Koltun, and S.~H. Chan, ``Dynamic low-light
  imaging with quanta image sensors,'' in {\em Proceedings of 16th European
  Conference on Computer Vision, Part XVII 16}, pp.~122--138, Springer, 2020.

\bibitem{gnanasambandam2019megapixel}
A.~Gnanasambandam, O.~Elgendy, J.~Ma, and S.~H. Chan, ``Megapixel
  photon-counting color imaging using quanta image sensor,'' {\em Optics
  Express}, vol.~27, no.~12, pp.~17298--17310, 2019.

\bibitem{li2021photon}
C.~Li, X.~Qu, A.~Gnanasambandam, O.~A. Elgendy, J.~Ma, and S.~H. Chan,
  ``Photon-limited object detection using non-local feature matching and
  knowledge distillation,'' in {\em Proceedings of the IEEE/CVF International
  Conference on Computer Vision}, pp.~3976--3987, 2021.

\bibitem{purelet}
J.~Li, F.~Luisier, and T.~Blu, ``Pure-let image deconvolution,'' {\em IEEE
  Transactions on Image Processing}, vol.~27, no.~1, pp.~92--105, 2017.

\bibitem{richardson1972bayesian}
W.~H. Richardson, ``Bayesian-based iterative method of image restoration,''
  {\em J. Opt. Soc. Am.}, vol.~62, no.~1, pp.~55--59, 1972.

\bibitem{lucy1974iterative}
L.~B. Lucy, ``An iterative technique for the rectification of observed
  distributions,'' {\em The Astronomical Journal}, vol.~79, p.~745, 1974.

\bibitem{venkatakrishnan2013plug}
S.~V. Venkatakrishnan, C.~A. Bouman, and B.~Wohlberg, ``Plug-and-play priors
  for model based reconstruction,'' in {\em 2013 IEEE Global Conference on
  Signal and Information Processing}, pp.~945--948, IEEE, 2013.

\bibitem{chan2016plug}
S.~H. Chan, X.~Wang, and O.~A. Elgendy, ``Plug-and-play {ADMM} for image
  restoration: Fixed-point convergence and applications,'' {\em IEEE
  Transactions on Computational Imaging}, vol.~3, no.~1, pp.~84--98, 2016.

\bibitem{rond2016poisson}
A.~Rond, R.~Giryes, and M.~Elad, ``{P}oisson inverse problems by the
  plug-and-play scheme,'' {\em Journal of Visual Communication and Image
  Representation}, vol.~41, pp.~96--108, 2016.

\bibitem{deblur_review}
M.~Bertero, P.~Boccacci, G.~Desider{\`a}, and G.~Vicidomini, ``Image deblurring
  with {P}oisson data: from cells to galaxies,'' {\em Inverse Problems},
  vol.~25, no.~12, p.~123006, 2009.

\bibitem{shepp1982maximum}
L.~A. Shepp and Y.~Vardi, ``Maximum likelihood reconstruction for emission
  tomography,'' {\em IEEE Transactions on Medical Imaging}, vol.~1, no.~2,
  pp.~113--122, 1982.

\bibitem{dey2006richardson}
N.~Dey, L.~Blanc-Feraud, C.~Zimmer, P.~Roux, Z.~Kam, J.-C. Olivo-Marin, and
  J.~Zerubia, ``Richardson--lucy algorithm with total variation regularization
  for 3d confocal microscope deconvolution,'' {\em Microscopy Research and
  Technique}, vol.~69, no.~4, pp.~260--266, 2006.

\bibitem{laasmaa2011application}
M.~Laasmaa, M.~Vendelin, and P.~Peterson, ``Application of regularized
  richardson--lucy algorithm for deconvolution of confocal microscopy images,''
  {\em Journal of Microscopy}, vol.~243, no.~2, pp.~124--140, 2011.

\bibitem{figueiredo2010restoration}
M.~A. Figueiredo and J.~M. Bioucas-Dias, ``Restoration of {P}oissonian images
  using alternating direction optimization,'' {\em IEEE Transactions on Image
  Processing}, vol.~19, no.~12, pp.~3133--3145, 2010.

\bibitem{harmany2011spiral}
Z.~T. Harmany, R.~F. Marcia, and R.~M. Willett, ``This is {SPIRAL}-{TAP}:
  Sparse {P}oisson intensity reconstruction algorithms—theory and practice,''
  {\em IEEE Transactions on Image Processing}, vol.~21, no.~3, pp.~1084--1096,
  2011.

\bibitem{nowak2000statistical}
R.~D. Nowak and E.~D. Kolaczyk, ``A statistical multiscale framework for
  {P}oisson inverse problems,'' {\em IEEE Transactions on Information Theory},
  vol.~46, no.~5, pp.~1811--1825, 2000.

\bibitem{blu2007sure}
T.~Blu and F.~Luisier, ``The {SURE}-{LET} approach to image denoising,'' {\em
  IEEE Transactions on Image Processing}, vol.~16, no.~11, pp.~2778--2786,
  2007.

\bibitem{xue2013multi}
F.~Xue, F.~Luisier, and T.~Blu, ``Multi-{W}iener {SURE}-{LET} deconvolution,''
  {\em IEEE Transactions on Image Processing}, vol.~22, no.~5, pp.~1954--1968,
  2013.

\bibitem{anscombe1948transformation}
F.~J. Anscombe, ``The transformation of {P}oisson, binomial and
  negative-binomial data,'' {\em Biometrika}, vol.~35, no.~3/4, pp.~246--254,
  1948.

\bibitem{makitalo2010optimal}
M.~Makitalo and A.~Foi, ``Optimal inversion of the {A}nscombe transformation in
  low-count {P}oisson image denoising,'' {\em IEEE transactions on Image
  Processing}, vol.~20, no.~1, pp.~99--109, 2010.

\bibitem{luisier2010fast}
F.~Luisier, C.~Vonesch, T.~Blu, and M.~Unser, ``Fast interscale wavelet
  denoising of {P}oisson-corrupted images,'' {\em Signal Processing}, vol.~90,
  no.~2, pp.~415--427, 2010.

\bibitem{zhang2008wavelets}
B.~Zhang, J.~M. Fadili, and J.-L. Starck, ``Wavelets, ridgelets, and curvelets
  for {P}oisson noise removal,'' {\em IEEE Transactions on Image Processing},
  vol.~17, no.~7, pp.~1093--1108, 2008.

\bibitem{azzari2016variance}
L.~Azzari and A.~Foi, ``Variance stabilization for noisy+ estimate combination
  in iterative {P}oisson denoising,'' {\em IEEE signal processing letters},
  vol.~23, no.~8, pp.~1086--1090, 2016.

\bibitem{sreehari2016plug}
S.~Sreehari, S.~V. Venkatakrishnan, B.~Wohlberg, G.~T. Buzzard, L.~F. Drummy,
  J.~P. Simmons, and C.~A. Bouman, ``Plug-and-play priors for bright field
  electron tomography and sparse interpolation,'' {\em IEEE Transactions on
  Computational Imaging}, vol.~2, no.~4, pp.~408--423, 2016.

\bibitem{ahmad2019plug}
R.~Ahmad, C.~A. Bouman, G.~T. Buzzard, S.~Chan, S.~Liu, E.~T. Reehorst, and
  P.~Schniter, ``Plug-and-play methods for magnetic resonance imaging: Using
  denoisers for image recovery,'' {\em IEEE signal processing magazine},
  vol.~37, no.~1, pp.~105--116, 2020.

\bibitem{dpir}
K.~Zhang, W.~Zuo, S.~Gu, and L.~Zhang, ``Learning deep {CNN} denoiser prior for
  image restoration,'' in {\em Proceedings of the IEEE Conference on Computer
  Vision and Pattern Recognition}, pp.~3929--3938, 2017.

\bibitem{he2019plug}
T.~He, Y.~Sun, B.~Chen, J.~Qi, W.~Liu, and J.~Hu, ``Plug-and-play inertial
  forward--backward algorithm for {P}oisson image deconvolution,'' {\em Journal
  of Electronic Imaging}, vol.~28, no.~4, p.~043020, 2019.

\bibitem{kamilov2017plug}
U.~S. Kamilov, H.~Mansour, and B.~Wohlberg, ``A plug-and-play priors approach
  for solving nonlinear imaging inverse problems,'' {\em IEEE Signal Processing
  Letters}, vol.~24, no.~12, pp.~1872--1876, 2017.

\bibitem{online_pnp}
Y.~Sun, B.~Wohlberg, and U.~S. Kamilov, ``An online plug-and-play algorithm for
  regularized image reconstruction,'' {\em IEEE Transactions on Computational
  Imaging}, vol.~5, no.~3, pp.~395--408, 2019.

\bibitem{buzzard2018plug}
G.~T. Buzzard, S.~H. Chan, S.~Sreehari, and C.~A. Bouman, ``Plug-and-play
  unplugged: Optimization-free reconstruction using consensus equilibrium,''
  {\em SIAM Journal on Imaging Sciences}, vol.~11, no.~3, pp.~2001--2020, 2018.

\bibitem{chan2019performance}
S.~H. Chan, ``Performance analysis of plug-and-play {ADMM}: A graph signal
  processing perspective,'' {\em IEEE Transactions on Computational Imaging},
  vol.~5, no.~2, pp.~274--286, 2019.

\bibitem{ryu2019plug}
E.~Ryu, J.~Liu, S.~Wang, X.~Chen, Z.~Wang, and W.~Yin, ``Plug-and-play methods
  provably converge with properly trained denoisers,'' in {\em International
  Conference on Machine Learning}, pp.~5546--5557, 2019.

\bibitem{romano2017little}
Y.~Romano, M.~Elad, and P.~Milanfar, ``The little engine that could:
  Regularization by denoising ({RED}),'' {\em SIAM Journal on Imaging
  Sciences}, vol.~10, no.~4, pp.~1804--1844, 2017.

\bibitem{cohen2020regularization}
R.~Cohen, M.~Elad, and P.~Milanfar, ``Regularization by denoising via
  fixed-point projection ({RED}-{PRO}),'' {\em SIAM Journal on Imaging
  Sciences}, vol.~14, no.~3, pp.~1374--1406, 2021.

\bibitem{reehorst2018regularization}
E.~T. Reehorst and P.~Schniter, ``Regularization by denoising: Clarifications
  and new interpretations,'' {\em IEEE Transactions on Computational Imaging},
  vol.~5, no.~1, pp.~52--67, 2018.

\bibitem{gregor2010learning}
K.~Gregor and Y.~LeCun, ``Learning fast approximations of sparse coding,'' in
  {\em Proceedings of the 27th International Conference on Machine Learning},
  pp.~399--406, 2010.

\bibitem{usrnet}
K.~Zhang, L.~V. Gool, and R.~Timofte, ``Deep unfolding network for image
  super-resolution,'' in {\em Proceedings of the IEEE/CVF Conference on
  Computer Vision and Pattern Recognition}, pp.~3217--3226, 2020.

\bibitem{li2019algorithm}
Y.~Li, M.~Tofighi, V.~Monga, and Y.~C. Eldar, ``An algorithm unrolling approach
  to deep image deblurring,'' in {\em ICASSP 2019 IEEE International Conference
  on Acoustics, Speech and Signal Processing (ICASSP)}, pp.~7675--7679, IEEE,
  2019.

\bibitem{li2020efficient}
Y.~Li, M.~Tofighi, J.~Geng, V.~Monga, and Y.~C. Eldar, ``Efficient and
  interpretable deep blind image deblurring via algorithm unrolling,'' {\em
  IEEE Transactions on Computational Imaging}, vol.~6, pp.~666--681, 2020.

\bibitem{yang2016deep}
Y.~Yang, J.~Sun, H.~Li, and Z.~Xu, ``Deep {ADMM}-net for compressive sensing
  {MRI},'' in {\em Advances in Neural Information Processing Systems},
  pp.~10--18, 2016.

\bibitem{padnet2019}
R.~Liu, S.~Cheng, L.~Ma, X.~Fan, and Z.~Luo, ``Deep proximal unrolling:
  Algorithmic framework, convergence analysis and applications,'' {\em IEEE
  Transactions on Image Processing}, vol.~28, no.~10, pp.~5013--5026, 2019.

\bibitem{monga2021algorithm}
V.~Monga, Y.~Li, and Y.~C. Eldar, ``Algorithm unrolling: Interpretable,
  efficient deep learning for signal and image processing,'' {\em IEEE Signal
  Processing Magazine}, vol.~38, no.~2, pp.~18--44, 2021.

\bibitem{gilton2021deep}
D.~Gilton, G.~Ongie, and R.~Willett, ``Deep equilibrium architectures for
  inverse problems in imaging,'' {\em IEEE Transactions on Computational
  Imaging}, vol.~7, pp.~1123--1133, 2021.

\bibitem{unet}
O.~Ronneberger, P.~Fischer, and T.~Brox, ``U-net: Convolutional networks for
  biomedical image segmentation,'' in {\em International Conference on Medical
  Image Computing and Computer-Assisted Intervention}, pp.~234--241, Springer,
  2015.

\bibitem{admm}
S.~Boyd, N.~Parikh, and E.~Chu, {\em Distributed Optimization and Statistical
  Learning Via the Alternating Direction Method of Multipliers}.
\newblock Now Publishers Inc, 2011.

\bibitem{lbfgs}
R.~H. Byrd, P.~Lu, J.~Nocedal, and C.~Zhu, ``A limited memory algorithm for
  bound constrained optimization,'' {\em SIAM Journal on Scientific Computing},
  vol.~16, no.~5, pp.~1190--1208, 1995.

\bibitem{figueiredo2009deconvolution}
M.~A. Figueiredo and J.~M. Bioucas-Dias, ``Deconvolution of {P}oissonian images
  using variable splitting and augmented lagrangian optimization,'' in {\em
  2009 IEEE/SP 15th Workshop on Statistical Signal Processing}, pp.~733--736,
  IEEE, 2009.

\bibitem{Flickr2K}
B.~Lim, S.~Son, H.~Kim, S.~Nah, and K.~Mu~Lee, ``Enhanced deep residual
  networks for single image super-resolution,'' in {\em Proceedings of the IEEE
  Conference on Computer Vision and Pattern Recognition Workshops},
  pp.~136--144, 2017.

\bibitem{motion_blur}
G.~Boracchi and A.~Foi, ``Modeling the performance of image restoration from
  motion blur,'' {\em IEEE Transactions on Image Processing}, vol.~21, no.~8,
  pp.~3502--3517, 2012.

\bibitem{levin}
A.~Levin, Y.~Weiss, F.~Durand, and W.~T. Freeman, ``Understanding and
  evaluating blind deconvolution algorithms,'' in {\em 2009 IEEE Conference on
  Computer Vision and Pattern Recognition}, pp.~1964--1971, IEEE, 2009.

\bibitem{kingma2014adam}
D.~P. Kingma and J.~Ba, ``Adam: A method for stochastic optimization,'' in {\em
  International Conference on Learning Representations (Poster)}, 2015.
\newblock Available online at \url{https://arxiv.org/abs/1412.6980} Accessed:
  Sep-20-2022.

\bibitem{xavier}
X.~Glorot and Y.~Bengio, ``Understanding the difficulty of training deep
  feedforward neural networks,'' in {\em Proceedings of the Thirteenth
  International Conference on Artificial Intelligence and Statistics},
  pp.~249--256, 2010.

\bibitem{diamond2017unrolled}
S.~Diamond, V.~Sitzmann, F.~Heide, and G.~Wetzstein, ``Unrolled optimization
  with deep priors,'' {\em arXiv preprint arXiv:1705.08041}, 2017.
\newblock Accessed: Sep-20-2022.

\bibitem{bsds300}
D.~Martin, C.~Fowlkes, D.~Tal, and J.~Malik, ``A database of human segmented
  natural images and its application to evaluating segmentation algorithms and
  measuring ecological statistics,'' in {\em Proceedings Eighth IEEE
  International Conference on Computer Vision. ICCV 2001}, vol.~2,
  pp.~416--423, IEEE, 2001.

\bibitem{dncnn}
K.~Zhang, W.~Zuo, Y.~Chen, D.~Meng, and L.~Zhang, ``Beyond a gaussian denoiser:
  Residual learning of deep {CNN} for image denoising,'' {\em IEEE Transactions
  on Image Processing}, vol.~26, no.~7, pp.~3142--3155, 2017.

\bibitem{elgendy2021low}
O.~A. Elgendy, A.~Gnanasambandam, S.~H. Chan, and J.~Ma, ``Low-light
  demosaicking and denoising for small pixels using learned frequency
  selection,'' {\em IEEE Transactions on Computational Imaging}, vol.~7,
  pp.~137--150, 2021.

\bibitem{photonstophotos}
``{Photons to Photos}.'' \url{https://www.photonstophotos.net/}.
\newblock Accessed: Sep-9-2021.

\end{thebibliography}


\begin{thebibliography}{1}

\bibitem{dpir}
K.~Zhang, W.~Zuo, S.~Gu, and L.~Zhang, ``Learning deep {CNN} denoiser prior for
  image restoration,'' in {\em Proceedings of the IEEE Conference on Computer
  Vision and Pattern Recognition}, pp.~3929--3938, 2017.

\bibitem{chan2016plug}
S.~H. Chan, X.~Wang, and O.~A. Elgendy, ``Plug-and-play {ADMM} for image
  restoration: Fixed-point convergence and applications,'' {\em IEEE
  Transactions on Computational Imaging}, vol.~3, no.~1, pp.~84--98, 2016.

\bibitem{grayworld}
G.~Buchsbaum, ``A spatial processor model for object colour perception,'' {\em
  Journal of the Franklin institute}, vol.~310, no.~1, pp.~1--26, 1980.

\end{thebibliography}
\bibliographystyle{ieeetr}

\end{document}


\title{Photon Limited Non-Blind Deblurring Using
Algorithm Unrolling -- Supplementary Material}

\author{Yash~Sanghvi,~\IEEEmembership{Student~Member,~IEEE}, Abhiram~Gnanasambandam,~\IEEEmembership{Student~Member,~IEEE}, and~Stanley~H.~Chan,~\IEEEmembership{Senior~Member,~IEEE}
\thanks{Y.~Sanghvi, A.~Gnanasambandam and S.~Chan are with the School of Electrical and Computer
Engineering, Purdue University, West Lafayette, IN 47907, USA. Email: {
\{ysanghvi, agnanasa, stanchan\}}@purdue.edu.} 
}
\maketitle

\section{ResUNet Architectural Details}
\begin{figure*}[!ht]
    \centering
    \includegraphics[page=1,width=0.99\linewidth]{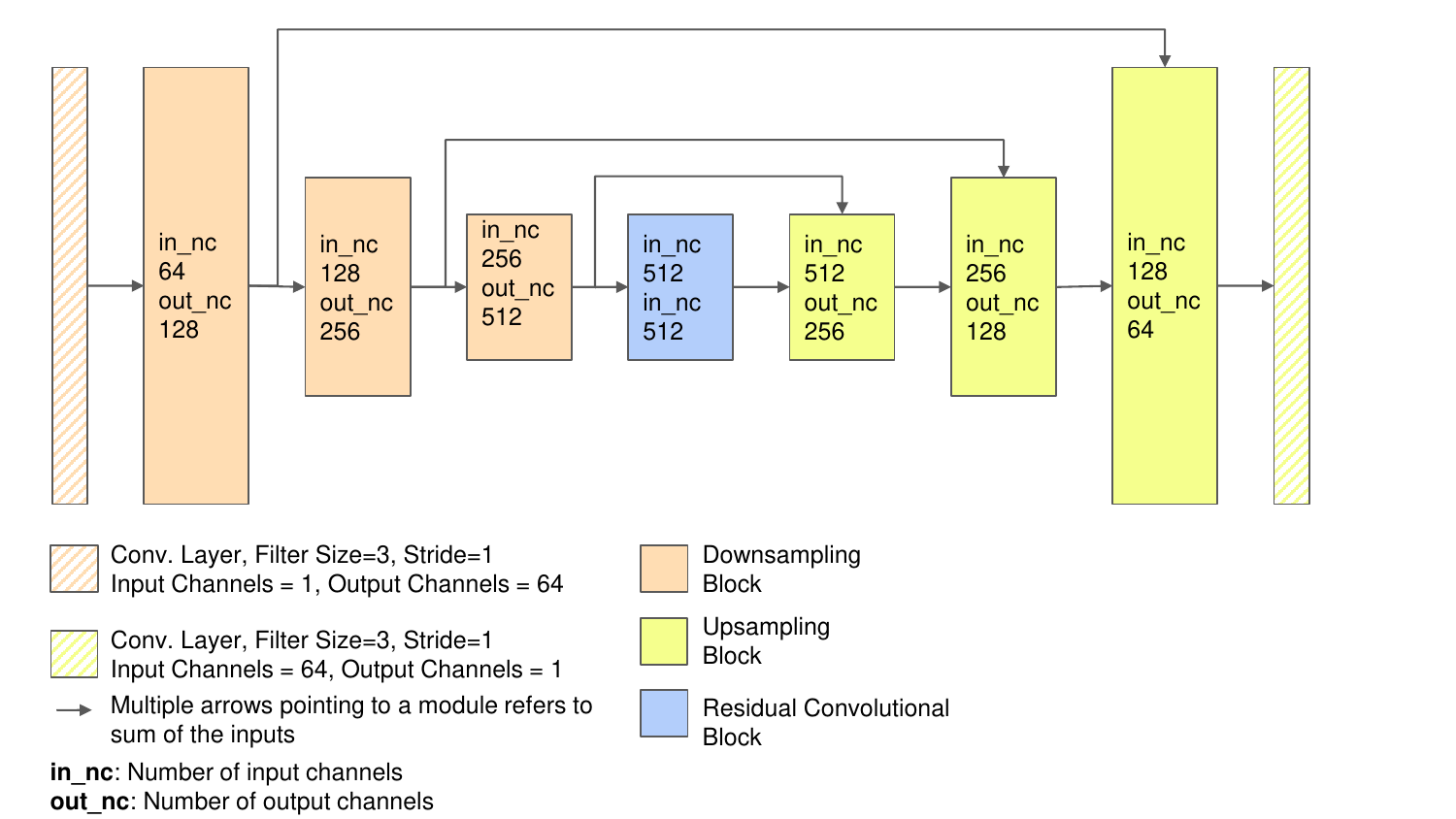}
    \caption{ \textbf{Architecture Details of ResUNet} - a version of the U-Net with skip connections as described in \cite{dpir}. Each downsampling block reduces the size of features by a factor of 2 and each upsampling block increases the size of features by the same factor. Note that the skip connections in ResU-Net are not concatenated to the other input going into a module but are added to them instead.}
    \label{fig:resunet}
\end{figure*}

In this section, we describe the architectural details of \emph{ResUNet} - the denoiser used in the proposed scheme. The ResUNet has an encoder-decoder structure with 4 downsampling, 1 residual convolutional and 4 upsampling blocks. The upsampling and downsampling blocks increases/decreases the size of the input by a factor of $\times 2$ and the residual blocks keeps the size of the image constant. The downsampling and upsampling blocks as shown in \fref{fig:resunet} are preceded and succeded by a convolutional layer each converting the image space into feature space and vice-versa. Note that except for the strided and transposed convolution filters, appropriate padding is added to each image to keep the output images size same as the input.

\textbf{Downsampling Block}: Each downsampling block consists of 2 residual sub-blocks followed by strided convolutional filter which downsizes the image size by factor of 2 but also doubles the number of output features. Each residual convolutional block has a convolutional layer, followed by ReLU activation followed by another convolutional layer. The output of these three layers is added to the original input - hence the term residual sub-block.

\textbf{Upsampling Block}: The upsampling block is similar to the downsampling block as described above. It consists of, in the sequence given, a transposed convolutional filter followed by 2 residuals sub-blocks. The strided convolutional filter upsamples the input by a factor $\times 2$ and halves the number of output features. Similar to those in the downsampling block, the residual sub-blocks keep the output size same as that of the output.

\textbf{Residual Convolutional Block}: The residual convolutional blocks consist of 3 resiudal sub-blocks in series with each other. In the ResUNet, it is found after the 4 downsampling operators and before the 4 upsampling operators.
\begin{figure*}[!ht]
    \centering
    \includegraphics[page=2,width=0.99\linewidth]{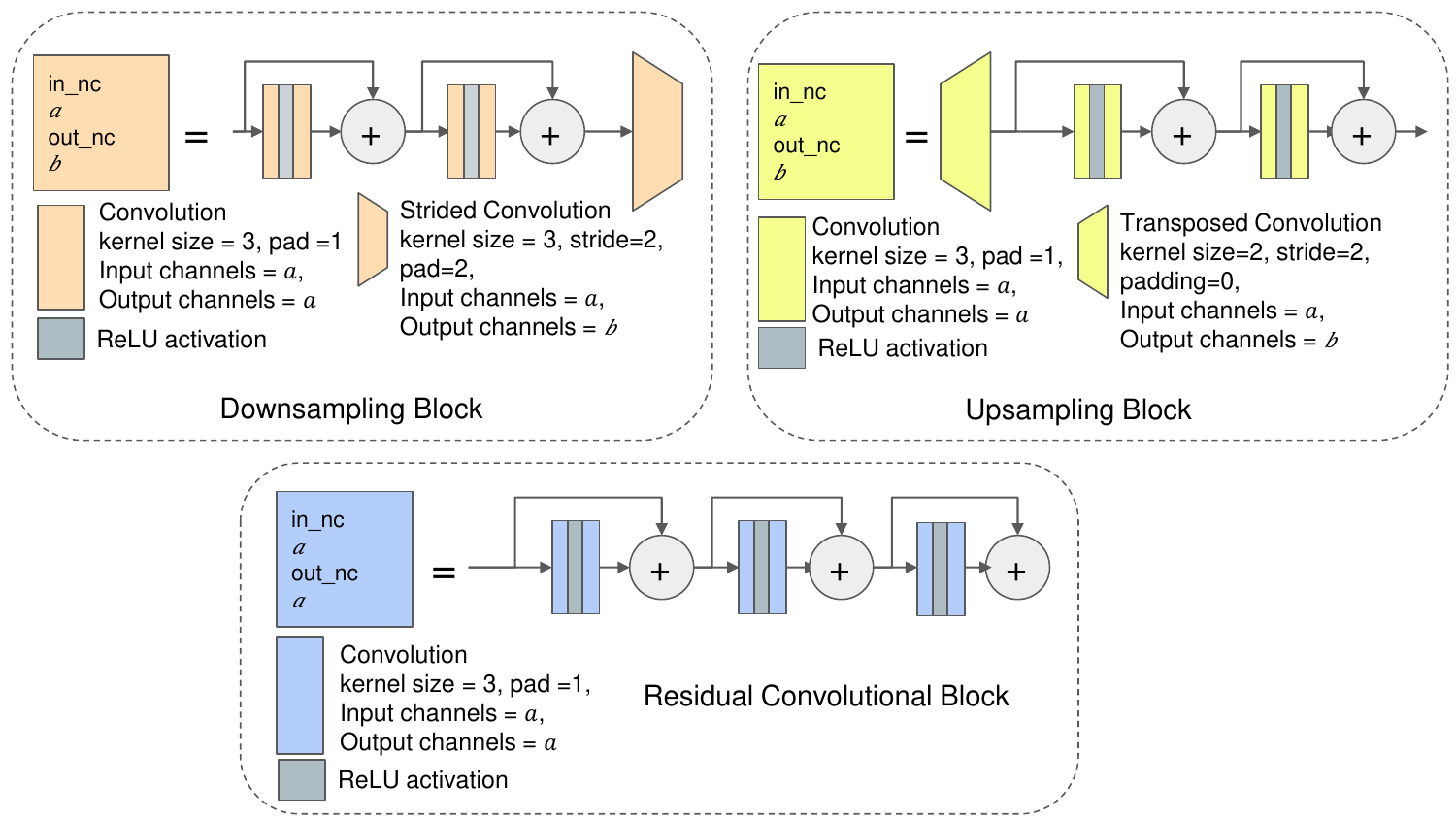}
    \caption{\textbf{ResUNet Blocks}. Description of various blocks shown in \fref{fig:resunet}. Note that all convolutional layers don't have a bias term in them.}
    \label{fig:blocks_resunet}
\end{figure*}

\section{Initialization of Hyperparameters}
Given the number of fixed iterations $K$, for which the scheme is unrolled, the hyperparameters $\rho_1^{k}, \rho_2^{k}$ for $k = 1,2,3,..., K$ are initialized using the photon level $\alpha$ and the blur kernel $\vk$ using the network \emph{InitNet}. In this section, we describe its architecture in detail.

The first stage consists of converting the blur kernel to a feature vector. First we ensure the kernel is of size $128 \times 128$. If not, then it is padded around the edges to ensure that. Then, we take the magnitude square of the spectrum of the kernel $\vk$, which is of the same size as the kernel i.e. $128 \times 128$. Then the spectrum-squared is used as the actual input to the convolutional block of the InitNet.

\begin{figure*}[!ht]
    \centering
    \includegraphics[page=3,width=0.99\linewidth]{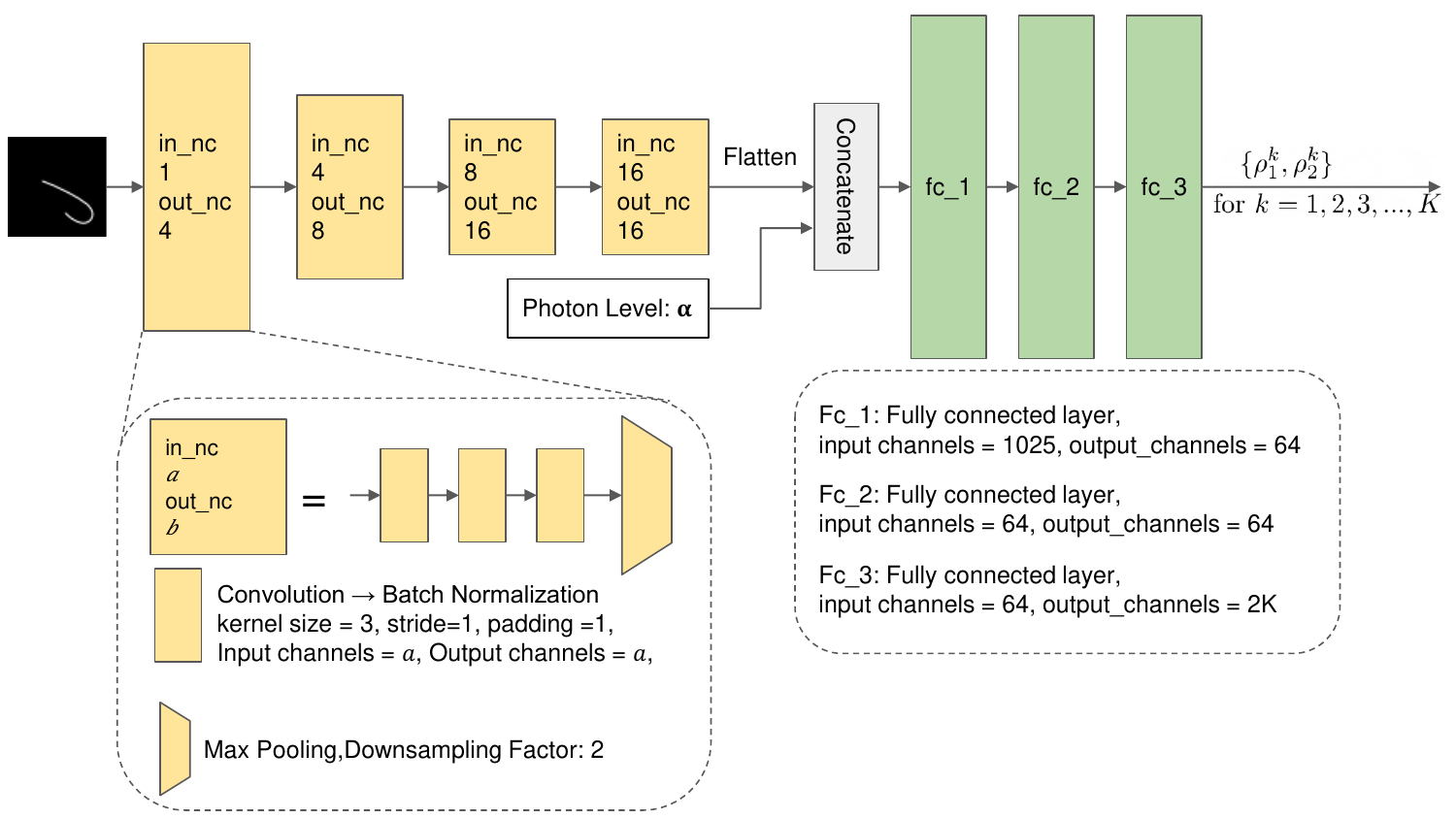}
    \caption{\textbf{Hyperparameter Network}. Description of network architecture used to determine the hyperparameters $\rho_1^{k}, \rho_2^{k}$ for $k = 1,2,...,K$}
    \label{fig:hyperparameters}
\end{figure*}

The convolutional block consists of 4 consecutive downsampling blocks (different from that of ResUNet). The details of the downsampling blocks are provided in \fref{fig:hyperparameters} and the output is half the size of the input image. After the downsampling blocks the output, of size $8 \times 16 \times 16$ is then flattened to a vector, concatenated with a scalar $\alpha$ to form a vector of size $2049 \times 1$. This is used as input the the fully connected part of the InitialNet (also described in \fref{fig:hyperparameters} )

\section{Comparison with Conventional Plug-and-Play}
In this section, we discuss the details of the experiment conducted in Section IV E where we compare the performance of conventional and alternate formulation of Plug-and-Play scheme. We evaluate the performance of the 3 schemes as described in the subsection on the \emph{BSD100} dataset. Each image in the dataset is randomly blurred by a kernel from Figure 6 from the main document a fixed photon level $\alpha$. The schemes are then separately evaluated at photon levels $5$, $10$, $20$, and $40$.

For the conventional PnP, we use the pretrained FFD-Net. Since the denoiser used is blind (i.e. agnostic to noise level), the output of the scheme is independent of $\sigma$ and dependent only on $\rho$.  We use the adaptive-update rule in \cite{chan2016plug} to update $\rho$ for each iteration. Specifically the following update rule is applied:
\begin{enumerate}
    \item if $\Delta^{(k)} > 0.99 \Delta^{(k-1)}$ then $\rho^{(k+1)} = 1.01\rho^{(k)}$
    \item if $\Delta^{(k)} \leq 0.99 \Delta^{(k-1)}$ then $\rho^{(k+1)} = \rho^{(k)}$
\end{enumerate}
where $\Delta^{(k)}$ is defined as follows
\begin{align}
\begin{split}
    \Delta^{(k)} \bydef  \frac{1}{3}\Big( ||\vx^{(k)}-\vx^{(k-1)}|| + \;||\vz^{(k)}-\vz^{(k-1)}|| \\ + \;||\vu^{(k)}-\vu^{(k-1)}||\Big)
    \end{split}
\end{align}
We terminate the PnP scheme when $\Delta_k$ fall below $10^{-2}$ or after 150 iterations, whichever happens first.

For the $\vx$-update step which needs to be solved as convex optimization solver, we use the memory-limited BFGS method. We use the implementation provided in \texttt{scipy.optimize} which requires the function to be optimized and the corresponding gradient value as its arguments. We use the default setting of the optimizer where it converges when the gradient norm is below $10^{-5}$.

The alternate formulation using 3-operator splitting is implemented in a similar manner as that of conventional PnP. The parameters $\rho_1^{(k)}, \rho_2^{(k)}$ determine the output of the scheme and are set to be same as each other  across iterations i.e. $\rho_1^{(k)} =  \rho_2^{(k)}= \rho_0^{(k)}$. $\rho_0^{(k)}$ is also adaptively updated using the following scheme
\begin{enumerate}
    \item if $\Delta^{(k)} > 0.99 \Delta^{(k-1)}$ then $\rho^{(k+1)} = 1.01\rho^{(k)}$
    \item if $\Delta^{(k)} \leq 0.99 \Delta^{(k-1)}$ then $\rho^{(k+1)} = \rho^{(k)}$
\end{enumerate}
where $\Delta^{(k)}$ is defined as follows
\begin{align}
\begin{split}
    \Delta^{(k)} \bydef  \frac{1}{5}\Big( ||\vx^{(k)}-\vx^{(k-1)}|| + \;||\vz^{(k)}-\vz^{(k-1)}|| \\+ \;||\vv_1^{(k)}-\vv_1^{(k-1)}||  +\;||\vu_1^{(k)}-\vu_1^{(k-1)}||+\\ \;||\vu_2^{(k)}-\vu_2^{(k-1)}||\Big)
    \end{split}
\end{align}

For a fixed photon level $\alpha$, we use the same $\rho^{(0)}$ and $\rho_0^{(k)}$ for 2-operator and 3-operator scheme respectively. Through trial-and-error, the initial $\rho$ values are varied as shown in Table I.
\begin{table}[!h]
    \label{tab:rho_val}
    \centering
     \caption{}
    \resizebox{0.5\linewidth}{!}{
    \begin{tabular}{c|c}
        Photon Level $\alpha$ & $\rho^{(0)}$/ $\rho_0^{(0)}$ \\
        \hline
        5 & 200 \\
        10 & 400 \\
        20 & 800 \\
        40 & 1000
    \end{tabular}}

\end{table}
\begin{figure}[H]
    \centering
    \includegraphics[trim={30 10 30 50},clip,width=0.9\linewidth]{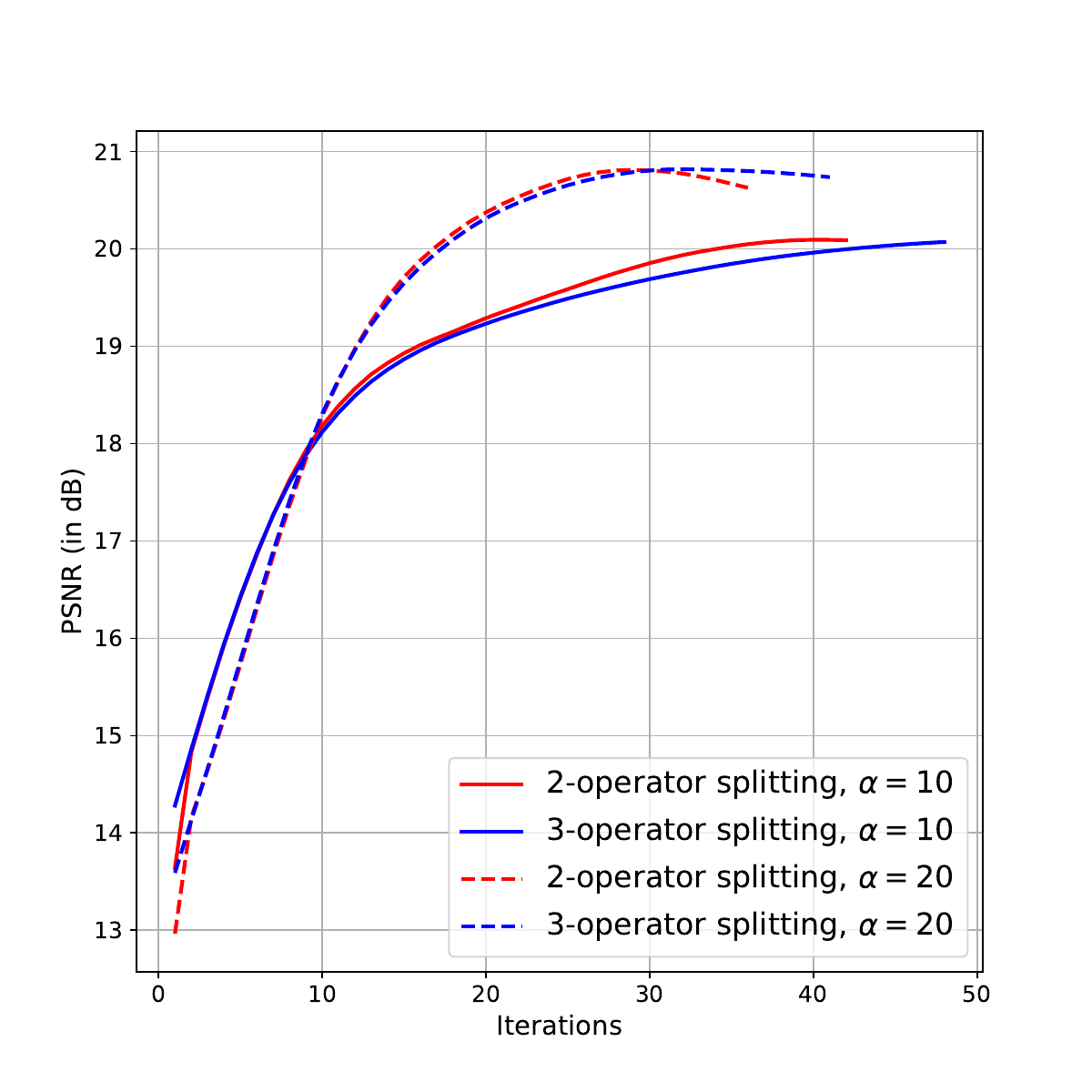}
    \caption{\textbf{Convergence of 2-operator splitting vs 3-operator splitting}.}
    \label{fig:convergence_3and2operator}
\end{figure}

\section{Reconstruction Implementation Details}
To reconstruct blurry and noisy real images, there are certain implementation details which need to be accounted for some assumptions for simulating Poisson blur are not valid for the real-world data.

One such example is the boundary conditions for convolution. We assume circular boundary conditions while blurring the image with a kernel since it allows us to implement convolution faster using FFT and provides a diagonalizable convolutional matrix $\mH$. However, we observed artifacts in our reconstructions as the circular boundary conditions may not apply when the the opposite ends of the image are not similar. To fix this issue, we pad the input image to double its original size by reflecting along the edges and then pushing it through the network. The relevant center portion is then cropped out from the network output.

Another issue we encounter when reconstructing real-world data is that the scalar quantity $\alpha$ indicative of the level of photon shot noise is not provided. We estimate it from the raw data (in terms of electrons generated) using the following heuristic.
\begin{align}
    \hat{\alpha} = \frac{\sum\limits_{i=1}^N \vy_{i}}{\beta N}
\end{align}
i.e. the photon level $\alpha$ is estimated to be equal to the average photon-per-pixels divided by a constant factor $\beta < 1$. Using trial-and-error, we set the constant factor $\beta = 0.33$.

To convert the individual grayscale image reconstructions i.e. \textit{R, G1, G2, B} into a color reconstruction, we need to account for color balancing. For this purpose, we use the gray-world assumption \cite{grayworld}. We normalize each channel reconstruction so that mean of each channel is equal to 1. This is followed by a combining grayscale images to form an RGB image using a demosaicing process.

\typeout{}
\bibliography{ref}
\bibliographystyle{ieeetr}